\documentclass[referee]{aa} 
\usepackage{graphics,epsfig}
\begin{document}

            %
            %
\title{Abundances of Cu and Zn in metal-poor stars:
\thanks{Based on the spectra collected with the 1.93$-$m telescope
of Haute Provence Observatory} clues for Galaxy evolution}
\titlerunning{Cu and Zn abundances in metal-poor stars}
%
\author{T.V. Mishenina\inst{1}, V.V. Kovtyukh\inst{1},
 C. Soubiran\inst{2}, C.Travaglio\inst{3,4} and M. Busso\inst{4,5}}
\authorrunning{Mishenina et al.} %
\offprints{T.V. Mishenina}
\institute{Odessa Astronomical Observatory, and Isaac Newton
 Institute of
 Chile, Odessa Branch, Shevchenko Park, Odessa, 65014, Ukraine\\
email: tamar@deneb.odessa.ua, val@deneb.odessa.ua \and
Observatoire de Bordeaux, CNRS UNR 5804, BP 89, 33270
Floirac, France\\
email: soubiran@observ.u-bordeaux.fr \and Max-Planck-Institut
f\"ur Astrophysik, Karl-Schwarzschild-Strasse 1,
D-85741 Garching bei M\"unchen, Germany\\
email: claudia@mpa-garching.mpg.de \and Istituto Nazionale di Astrofisica
(INAF) - Osservatorio Astronomico di Torino, via Osservatorio 20,
10025 Pino Torinese (To), Italy
\and Department of Physics, University of Perugia, Italy\\
email: busso@pg.infn.it}
\date{Received,2002; accepted }
\abstract{We present new observations of copper and zinc
abundances in 90 metal-poor stars, belonging to the metallicity range
--3$<$[Fe/H]$<$--0.5. The present study is based on high resolution
spectroscopic measurements collected at the Haute Provence
Observatoire (R= 42000, S/N$>$100). The trend of Cu and Zn abundances as
a function of the metallicity [Fe/H] is discussed and compared to
that of other heavy elements beyond iron. We also estimate
spatial velocities and galactic orbital parameters for our
target stars in order to disentangle the population of
disk stars from that of halo stars using kinematic criteria.
In the absence of a firm a priori
knowledge of the nucleosynthesis mechanisms controlling
Cu and Zn production, and of the relative stellar sites, we
derive constraints on these last from the trend of the observed
ratios [Cu/Fe]and [Zn/Fe] throughout the history of
the Galaxy, as well as from a few well established properties of
basic nucleosynthesis processes in stars. We thus confirm
that the production of Cu and Zn requires a number of different
sources (neutron captures in massive stars, $s$-
processing in low and intermediate mass stars, explosive
nucleosynthesis in various supernova types). We also attempt a
ranking of the relative roles played by different production
mechanisms, and verify these hints through a simple estimate of
the galactic enrichment in Cu and Zn. In agreement with
suggestions presented earlier, we find evidence that
Type Ia Supernovae must play a relevant role, especially
for the production of Cu.
\keywords{Stars: abundances,
nucleosynthesis -- Stars: spectroscopy, kinematics: --
Galaxy: evolution}}
\maketitle
\section{Introduction}

The introduction of efficient high-resolution spectrographs on
modern telescopes has greatly enhanced our ability to reconstruct
the chemical enrichment of the Galaxy, which derives from a gradual
melding of the outcomes by various nucleosynthesis processes in
stars. This is obtained by the return of new elements into the
interstellar medium (ISM) through slow and fast mass loss
phenomena, whose relative importance is controlled by several
parameters, like the initial stellar mass distribution, the physics
of stellar winds, star formation rates, stellar lifetimes,  etc.
This picture is made even more complex by the different dynamical
behavior of the various galactic subsystems in a very
clumpy galactic structure, which both sets the
time scales of mixing processes in the ISM, and controls
the hierarchy of cluster and individual star formation. A galactic
evolution model describing all these processes in detail, including
the dynamical interactions between subsystems, is still lacking,
because of the enormous complexities inherent in such a construction
and of the high number of free parameters inevitably involved in it.
Unfortunately, for many heavy elements we infer that the mechanisms
through which their galactic enrichment was achieved do involve all
these ill-modeled complexities, so that for them galactic astrophysics
has reached a stage in which observations lead theory by a large
distance.

An example of this is provided by Cu and Zn, two elements
immediately following the iron peak, for which we present in
this paper new measurements for a very large sample (90) of
metal-poor stars. In our previous works (Mishenina \&
Kovtyuk~2001, hereafter Paper I; Mishenina et al.~2001, hereafter
Paper II) we investigated the behavior of several heavy elements
at different metallicities. The present study on copper and zinc
continues those efforts.

>From the observational point of view, Cu and Zn abundances were
early addressed by  Gratton \& Sneden~(1988) and by Sneden \&
Crocker~(1988). Their data favored a primary-like type of
production for Zn (i.e. mechanisms yielding constant enrichment
relative to iron) and a secondary-like process for Cu (i.e. one
requiring iron seeds from previous stellar generations, giving
rise to an enrichment proportional to the iron content). Later,
further abundances for the two elements in halo
and disk stars were provided by Sneden, Gratton, \& Crocker~1991,
and subsequently by Primas et al.~2000 and
Blake et al.~2001. Recently, also some data for $\omega$ Centauri
have become available (see e.g. Smith et al.~2000; Cunha et al.~2002;
Pancino et al.~2002), together with abundances in single very
low-metallicity galactic stars (Westin et al.~2000; Cowan et al.~2002;
Hill et al.~2002). Measurements of Zn in damped Ly-$\alpha$ systems (see
e.g. Pettini et al.~1999; Molaro et al.~2000) complete the list,
offering an opportunity to address the production of heavy
elements in the Universe at an epoch immediately following Galaxy
formation.

Despite those recent integrations, more than a decade has passed
since the work by Sneden et al.~(1991), who provided the only
large sample of Cu and Zn abundances in metal-poor stars of
our Galaxy. That work was used repeatedly to investigate the
nucleosynthesis of these elements. With the present study we aim
at providing a significant update of available database, and at
establishing more reliable constraints on the still open problems
involved in Cu and Zn formation.

The first schematic description of the chemical evolution of Cu
and Zn was proposed by Sneden et al.~(1991), who suggested that
they might be ascribed mainly to the weak $s$-process. Their
conclusions were subsequently questioned by Raiteri et al.~(1992)
and by Matteucci et al.~(1993). In this last work evidence was
presented in favor of a large contribution from relatively
long-lived processes, tentatively identified as Type Ia supernovae.
Contrary to this, Timmes et al.~(1995), using the copper and zinc
yields of Type II supernova explosion from Woosley \&
Weaver~(1995), suggested that these elements might be synthesized
in significant amounts by the major nuclear burning stages in
massive stars. These contrasting explanations are an example of
the large uncertainties one meets when an incomplete picture of
stellar yields and a simplified chemical evolution scheme have to
be used for interpreting the data. Simplified, and sometimes
purely analytical, chemical evolution models were very
useful in the past, before the advent of high resolution
spectroscopy (Lynden-Bell~1975; Tinsley \& Larson~1978;
Tinsley~1980; Clayton~1984). However, they are no longer
sufficient today, after more sophisticated and precise
measurements have become available.

Waiting for a revision in the nucleosynthesis models, what one can
do is to provide a homogeneous set of data, derived with
the same methods for many stars, and to compare them with known
results for other neutron-rich elements. This is actually the
scope of the present work, based on abundances obtained in a
homogeneous way for metal-poor stars belonging either to the halo
or to the thick disk of the Galaxy. Using the data as a
guideline, we shall examine which scenarios for the stellar
synthesis of these elements are compatible with the observed
trends. As a consistency check, we shall then verify our
conclusions through a simple computation of the ensuing
chemical enrichment of the Galaxy, making use of the model
adopted by Travaglio et al.~(1999).

\section{Observations and reduction}

The observational data for the sample stars were taken from the
library of spectra obtained with the ELODIE spectrograph on the
1.93m telescope of the Haute Provence Observatoire (Soubiran, Katz
\& Cayrel~1998). The resolution of spectra is R = 42,000, signal
to noise ratios are always larger than 100, and the spectral
region explored is between 4400 and 6700 \AA. A preliminary
reduction of these spectra was made by Katz et al.
(\cite{katzet98}). Further processing was done through the DECH20
software (Galazutdinov \cite{gal92}). A comparison of the
equivalent widths (EWs) measured by us with those from other
sources in the literature was previously given in Paper I and
Paper II, yielding a good agreement.

\subsection{Model atmosphere parameters}

The effective temperatures (T$_{eff}$), surface gravities ($\log$ g),
metallicities and microturbulence velocities (V$_{t}$) were
determined earlier by us (see Paper I). A comparison of our
parameters with those by other authors was discussed
in the same work, showing good agreement within the
uncertainties. T$_{eff}$ was derived by fitting the wing of the
H$_{\alpha}$ line in the observed and calculated spectra.
Synthetic spectra were computed with the STARSP code (Tsymbal
\cite{tsymbal96}) and with the grid of atmosphere models of
Kurucz (\cite{kurucz93}). The log g values were specified by
using the condition of ionization balance between Fe I and Fe II
lines. V$_{t}$ values were obtained by imposing independence from
the EW of the abundance derived from each line. The accuracy in
the estimate of the microturbulence velocity is $\pm$ 0.2 km/s;
corresponding estimates for T$_{eff}$ are $\pm$ 100 K; for $\log
g$ $\pm$ 0.3 dex. We note that we re-determined the value of
the micro-turbulence velocity for HD 122563, using only the lines
with EW $>$ 20 m\AA~,  which are not sensitive to V$_{t}$. The
new velocity is V$_{t}$ = 2.2 km/s. As a consequence, for this
star the values of [Fe/H] and of the Ba abundance, estimated with
the new value of V$_{t}$, supersede those previously presented.

\section{Copper and zinc abundances and their uncertainties}

As indicated previously, abundance measurements were carried out
using the atmosphere models by Kurucz (\cite{kurucz93}).
Appropriate models for each star were derived by means of standard
interpolation through T$_{\rm eff}$ and $\log$ g. The model
metallicities were taken with the accuracy $\pm$0.25 dex. See
below for comments on the analysis of each element.

\subsection{Copper}

The 5105.54, 5218.20, 5782.12 \AA~ lines of Cu I were used for
abundance definition. Synthetic spectra were calculated
taking into account the hyper-fine structure of Cu I components
(Steffen \cite{steffen85}). Fig.~1 shows a comparison
between the observed (dot) and computed (line) spectrum for HD 117876,
near the Cu I 5105.54 \AA line.

\begin{figure}
\resizebox{\hsize}{!}{\includegraphics{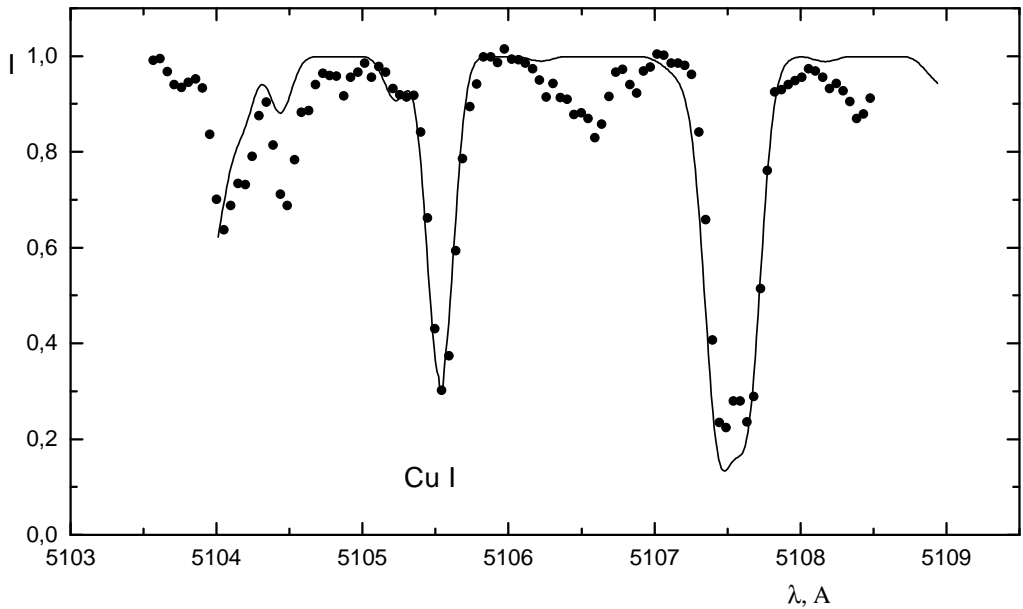}}
\caption[]{Comparison between observed {\it dot} and computed
{\it line} spectra for HD 117876 in the region of the line of Cu
I at 5105.54 \AA.}
\label{Fig1}
\end{figure}

\subsection{Zinc}

The abundance of zinc was obtained from the 4722.16, 4810.53,
6362.35 \AA~ lines of Zn I. We used the EWs of these lines and
the Kurucz's WIDTH9 model for the determination of the zinc
abundance. Oscillator strengths for Zn I lines were
taken from Gurtovenko \& Kostyk (\cite{gurko89}).

\subsection{Kinematics}

In order to distinguish disk stars (D) from halo stars (H) by
kinematic criteria, we determined the spatial velocities
and galactic orbital parameters of our target stars. Table 1 lists
the 3 components (U, V, W) of the spatial velocities with respect to
the LSR, and the parameters of the orbits: apo-galactic and
peri-galactic distance (R$_{max}$, R$_{min}$), maximum height
over the galactic plane (Z$_{max}$) and eccentricity ($e$).
The equations of motion were integrated in the galactic
model by Allen \& Santillan (\cite{allsan91}) over an age of 12
Gyr. The adopted velocity of the Sun with respect to the LSR is
(9, 5, 6) km/s, the solar galactocentric distance is assumed
to be R$_{\odot}$=8.5 kpc and its assumed circular velocity
is V$_{lsr}$=220 km/s.

\begin{table*}
\caption[]{Kinematic parameters for our program stars} \label{T1}
\begin{tabular}{lrrrrrrrrc}
\hline Star  &    [Fe/H] & U  &  V & W & R$_{min}$& R$_{max}$&
$|$Z$_{max}$$|$
& e & Pop \\
\hline \object{HD 245   }&--0.78& --29.6&--111.5& --49.7&+2.85&
+8.65&
+0.90&0.50&I\\
\object{HD 2796  }&--2.21&  +79.5&--108.1&  +44.2&+2.87& +9.35&
+1.02&0.53&I\\
\object{HD 3546  }&--0.63& +111.6& --54.8&   +3.3&+4.46&+10.77&
+0.05&0.41&D\\
\object{HD 3567  }&--1.20& +144.5&--230.8& --36.9&+0.16&+10.70&
+7.68&0.97&H\\
\object{HD 4306  }&--2.52& --43.1& --26.7&  +69.6&+6.68& +9.34&
+1.52&0.17&D\\
\object{HD 5395  }&--0.19&  +58.8& --20.6&  --5.0&+6.41& +9.69&
+0.06&0.20&D\\
\object{HD 5916  }&--0.51& --10.9& --94.1&  +10.6&+3.42& +8.60&
+0.14&0.43&D\\
\object{HD 6582  }&--0.89& --33.4&--154.4& --29.3&+1.39& +8.61&
+0.47&0.72&I\\
\object{HD 6755  }&--1.47&--123.5&--467.6&  +87.5&+6.95&+17.53&
+3.24&0.43&H\\
\object{HD 6833  }&--0.89& +130.2&--198.1&  +79.6&+0.40&+10.96&
+6.92&0.93&H\\
\object{HD 8724  }&--1.65&  +41.4&--177.5&  +13.9&+0.81& +8.84&
+0.45&0.83&H\\
\object{HD 10700 }&--0.56&  +27.5&  +34.0&  +18.9&+8.32&+11.98&
+0.28&0.18&D\\
\object{HD 13530 }&--0.48& --96.0& --47.7& --15.8&+4.82&+10.41&
+0.21&0.37&D\\
\object{HD 13783 }&--0.61&  +57.6&  +11.4& --71.5&+7.67&+11.70&
+1.72&0.21&D\\
\object{HD 15596 }&--0.67&  +87.4& --68.2&  +56.0&+4.31& +9.92&
+1.11&0.39&D\\
\object{HD 18768 }&--0.51& --79.4&  +43.5& --15.2&+7.65&+15.01&
+0.26&0.32&D\\
\object{HD 19445 }&--1.89& +165.4&--118.4& --61.3&+2.28&+12.19&
+1.55&0.69&I\\
\object{HD 23439 }&--1.14& --86.4&--109.9& --70.4&+2.85& +9.53&
+1.54&0.54&I\\
\object{HD 25329 }&--1.73& --31.9&--184.8&  +26.3&+0.63& +8.59&
+4.64&0.86&H\\
\object{HD 26297 }&--1.91& --16.7& --44.1&  +36.4&+5.89& +8.75&
+0.62&0.20&D\\
\object{HD 37828 }&--1.49&--101.7&--163.5& --46.6&+1.19& +9.83&
+1.71&0.78&I\\
\object{HD 44007 }&--1.49& --71.4&--158.1&  +17.8&+1.31& +9.14&
+0.46&0.75&I\\
\object{HD 45282 }&--1.28&--236.8&--181.4& --37.4&+0.64&+15.99&
+7.95&0.92&H\\
\object{HD 46480 }&--0.49&   +4.4& --49.9& --77.6&+5.87& +8.59&
+1.65&0.19&D\\
\object{HD 51530 }&--0.39&  --7.2&  +36.8& --16.8&+8.54&+12.04&
+0.25&0.17&D\\
\object{HD 63791 }&--1.67&  +16.7&--123.6&--100.5&+2.85& +8.77&
+3.45&0.51&I\\
\object{HD 64090 }&--1.69& +269.4&--215.7&
--81.9&+0.07&+20.02&+15.71&0.99&H\\
\object{HD 64606 }&--0.82& --68.5& --55.9&   +5.8&+4.75& +9.39&
+0.07&0.33&D\\
\object{HD 76932 }&--0.90& --39.1& --85.3&  +75.5&+3.99& +8.80&
+1.66&0.38&D\\
\object{HD 84937 }&--2.00& +234.7&--233.1&
--1.3&+0.29&+15.30&+12.19&0.96&H\\
\object{HD 87140 }&--1.71& +133.0& --11.0&  +60.3&+5.85&+14.18&
+1.74&0.42&D\\
\object{HD 88609 }&--2.66&  +34.5& --45.2&  --9.2&+5.62& +8.94&
+0.26&0.23&D\\
\object{HD 88725 }&--0.65&  +80.6& --28.3& --16.1&+5.73&+10.27&
+0.21&0.28&D\\
\object{HD 94028 }&--1.43& --26.5&--135.6&  +17.1&+1.96& +8.60&
+0.21&0.63&I\\
\object{HD 103095}&--1.39& +289.9&--154.1&  --6.9&+1.08&+22.01&
+0.21&0.91&H\\
\object{HD 105755}&--0.65&  +44.9&  --2.9& --19.4&+7.37& +9.82&
+0.27&0.14&D\\
\object{HD 108076}&--0.85& --84.9& --36.7&  --9.2&+5.36&+10.19&
+0.13&0.31&D\\
\object{HD 108317}&--2.17&--128.2&--106.3& --12.4&+2.70&+10.53&
+0.31&0.59&I\\
\object{HD 110184}&--2.27&   +9.7& --70.2& +130.6&+6.01& +8.67&
+4.10&0.18&I\\
\object{HD 114762}&--0.72& --73.8& --64.5&  +63.9&+4.55& +9.54&
+1.31&0.35&D\\
\hline
\end{tabular}
\end{table*}
%
\begin{table*}
{Table 1 (Continued)}\\
\begin{tabular}{lrrrrrrrrc}
\hline Star  &    [Fe/H]     & U  &  V & W & R$_{min}$&
R$_{max}$& $|$Z$_{max}$$|$
& e & Pop  \\
\hline \object{HD  117876}&--0.47& +125.3& --71.5&
+1.5&+3.76&+10.94&
+0.17&0.49&I\\
\object{HD  122563}&--2.66&--124.7&--212.4&  +19.1&+0.15& +9.92&
+7.10&0.97&H\\
\object{HD  122956}&--1.60&  +34.0&--186.0& +118.0&+0.93& +8.46&
+6.05&0.80&H\\
\object{HD  124897}&--0.58&  +34.1&--114.6&   +2.6&+2.62& +8.63&
+0.03&0.53&I\\
\object{HD  127243}&--0.65& --65.8&--100.0&  +55.5&+3.16& +9.12&
+1.02&0.49&I\\
\object{HD  132142}&--0.51& --99.7& --50.3&  +25.4&+4.70&+10.44&
+0.37&0.38&D\\
\object{HD  134169}&--0.72&  +24.9&   +2.8&  +20.4&+7.94& +9.35&
+0.27&0.08&D\\
\object{HD  140283}&--2.50&--239.9&--247.2&  +48.5&+0.45&+16.01&
+9.48&0.95&H\\
\object{HD  150177}&--0.64&   +2.7& --19.4& --19.2&+7.12& +8.47&
+0.23&0.09&D\\
\object{HD  157089}&--0.56&--157.9& --36.2&  --4.3&+4.55&+13.70&
+0.07&0.50&I\\
\object{HD  159482}&--0.86&--156.2& --57.8&  +86.5&+4.19&+13.49&
+2.79&0.53&I\\
\object{HD  160693}&--0.46& +213.8&--112.4&  +90.0&+2.55&+15.89&
+5.41&0.72&I\\
\object{HD  165195}&--2.03&  +76.3& --99.4& --10.8&+3.03& +8.94&
+0.14&0.49&I\\
\object{HD  165908}&--0.61&   +3.1&   +5.7&  +15.5&+8.48& +8.99&
+0.19&0.03&D\\
\object{HD  166161}&--1.20& +130.4&--126.7&   +8.3&+2.04&+10.26&
+0.11&0.67&I\\
\object{HD  175305}&--1.42& --60.4&
--77.4&--287.4&+8.18&+22.88&+19.82&0.47&H\\
\object{HD  184499}&--0.64& --55.9&--156.8&  +64.6&+1.35& +8.93&
+1.29&0.74&I\\
\object{HD  187111}&--1.74&--143.7&--154.8& --37.1&+1.16&+10.42&
+0.63&0.80&H\\
\object{HD  188510}&--1.48&--143.2&--108.6&  +68.6&+2.63&+11.31&
+1.74&0.62&I\\
\object{HD  189558}&--1.00&  +84.5&--122.5&  +49.2&+2.35& +9.28&
+1.01&0.60&I\\
\object{HD  194598}&--1.16& --66.8&--271.4& --24.5&+0.99& +8.92&
+0.46&0.80&H\\
\object{HD  195633}&--0.55& --49.2& --15.5&  --5.4&+6.61& +9.48&
+0.08&0.18&D\\
\object{HD  201889}&--0.85&--120.4& --77.3& --31.2&+3.59&+10.71&
+0.50&0.50&I\\
\object{HD  201891}&--0.99& +101.0&--110.1& --52.6&+2.71& +9.82&
+1.03&0.57&I\\
\object{HD  204155}&--0.78& --25.2&--120.5& --38.5&+2.46& +8.55&
+0.59&0.55&I\\
\object{HD  204543}&--1.79&   +6.3&--138.0&  +21.2&+1.85& +8.27&
+0.43&0.63&I\\
\object{HD  208906}&--0.71&  +82.2&   +2.9&  --5.0&+6.79&+11.55&
+0.07&0.26&D\\
\object{HD  216143}&--2.11& +138.7&--155.1&  +90.6&+1.67&+10.94&
+4.63&0.73&I\\
\object{HD  216174}&--0.56& --29.0& --40.7&   +5.7&+5.72& +8.78&
+0.07&0.21&D\\
\object{HD  218502}&--1.72&  +12.3& --99.5&  --2.7&+3.18& +8.50&
+0.07&0.46&I\\
\object{HD  218857}&--1.84&  +53.8&--124.5& +158.1&+3.98& +9.39&
+6.37&0.40&I\\
\object{HD  219617}&--1.43& +391.8&--325.5& --51.7&+1.61&+57.95&
+5.09&0.95&H\\
\object{HD  221170}&--2.05& +111.3&--127.6& --16.0&+2.25& +9.93&
+0.41&0.63&I\\
\object{HD  221377}&--0.88& --33.9&  +12.7& --23.4&+7.98&+10.40&
+0.33&0.13&D\\
\object{HD  224930}&--0.85&   +0.3& --66.3& --26.4&+4.60& +8.50&
+0.34&0.30&D\\
\object{HD  338529}&--2.31&  +38.3&--154.7& --58.8&+1.44& +8.71&
+1.01&0.72&I\\
\object{HD  345957}&--1.33& --20.8& --99.3& +106.1&+3.79& +8.54&
+3.18&0.39&D\\
\object{BD--18 5550}&--3.01& --69.3&--183.8&--20.7&+0.57& +8.63&
+4.81&0.88&H\\
\object{BD+02 3375}&--2.26&--357.5&--245.7&
+84.3&+0.85&+38.35&+14.53&0.96&H\\
\object{BD+02 4651}&--1.82&--220.7&--374.0&  +12.3&+3.40&+17.24&
+0.44&0.67&H\\
\object{BD+04 4551}&--1.51& --80.1& --49.0&  +78.1&+5.16&+10.02&
+1.87&0.32&D\\
\hline
\end{tabular}
\end{table*}
\begin{table*}
{Table 1 (Continued)}\\
\begin{tabular}{lrrrrrrrrc}
\hline Star  &    [Fe/H] &   U  &  V & W & R$_{min}$& R$_{max}$&
$|$Z$_{max}$$|$
& e & Pop \\
\hline \object{BD+17 4708}&--1.56&--293.0&--268.6&
+8.6&+0.80&+21.98&
+0.24&0.93&H\\
\object{BD+23 3130}&--2.62&--103.6&--430.6& +154.9&+6.90&+15.38&
+6.79&0.38&H\\
\object{BD+29 0366}&--1.01& --52.9& --70.7& --42.0&+4.32& +9.03&
+0.67&0.35&D\\
\object{BD+29 2091}&--1.93& +161.0&--339.4& +102.2&+3.19&+12.99&
+4.20&0.61&H\\
\object{BD+30 2611}&--1.41& --33.2&
--80.0&--264.6&+8.21&+17.22&+14.70&0.35&H\\
\object{BD+36 2165}&--1.51& +214.7&--255.5&--129.1&+0.56&+16.77&
+9.18&0.94&H\\
\object{BD+41 3931}&--1.68&  +80.2&--137.5& --87.9&+2.12& +9.42&
+2.10&0.63&I\\
\object{BD+42 3607}&--1.97&--164.6&--162.1&  +20.1&+1.03&+11.45&
+0.77&0.84&H\\
\object{BD+66 0268}&--1.95&--166.0&--414.5& --66.5&+4.97&+15.55&
+2.02&0.52&H\\
\hline
\end{tabular}
\begin{list}{}{}
\item --$Note$. The velocities U, V, W (in km/s) are given with
respect to the LSR, with U positive towards the GC; R$_{min}$,
$|$$R_{max}$$|$, Z$_{max}$ are in kpc.
\end{list}
\end{table*}

The first group of stars (D) includes objects with disk-like kinematics,
(nearly-circular orbits, not reaching extreme distances from the plane and
with a significant component of rotational velocity. This group is
dominated by thick-disk stars. We chose to include in category D
stars with height Z$_{max}<$ 4.0 kpc, eccentricity $e <$0.45 and rotational
velocity V$_{rot}>$ 160 km/s. The second group (H) includes all
stars whose orbits reach larger distances from the
galactic plane; they have large values of $e$ or retrograde
rotational velocities. The values of Z$_{max}>$6 kpc, or
e$>$0.75, or V$_{rot}<$ 0 characterize such cases. All other
stars fall into an intermediate group (I) corresponding to the
overlap between the thick disk and the halo; this includes 4
stars with [Fe/H]\footnotemark \footnotetext{In the classical
notation where [X/H]$=$ log(N$_X$/N$_H$)$_*$ -
log(N$_X$/N$_H$)$_\odot$} $<$ -1.6 (HD 4306, HD 26297, HD 87140,
HD 88609), but showing disk-like orbits. No star with parameters
typical of the thin galactic disk was selected for our work:
hence, kinematic data simply confirm the chemical evidence that
we are dealing with stars of the oldest subsystems in the Galaxy.

\subsection{Analysis of the abundance data}

As mentioned above, a detailed analysis of atmospheric
parameters, EWs and abundance results for some elements was
given in our previous works (Paper I, Paper II). The main sources
of uncertainty in the final element abundances lay in: 1)
spectrum processing (from flat-field division to measurement of
EWs); 2) atmosphere modeling (values of T$_{eff}$,
$\log$ g, V$_{t}$, [Fe/H]; thermal and dynamical structure
of the atmosphere; LTE or nLTE approach, etc.); 3) estimating
parameters from atomic physics (oscillator strengths, radiative
and collisional damping constants, hyper-fine structure etc.).

Here we shall consider uncertainties related to atmospheric
parameters and to spectrum processing (i.e those involved in the
measurement of EWs and in the fits with synthetic spectra,
including the uncertainty arising from the positioning of the
spectral continuum). At the end of this subsection we shall also
comment on the limits of the LTE abundance analysis as employed
here.

The influence of atmospheric parameters on the final abundances
can be estimated through the sensitivity of [Cu/Fe]
and [Zn/Fe] to variations in the model parameters. Table~2 lists
the results of such computations, assuming typical atmospheric
parameter uncertainties ($\Delta$ T$_{eff}$ = +100 K; f $\Delta$
$\log$ g= +0.3; $\Delta$ V$_{t}$=+0.2 km s$^{-1}$; $\Delta$[Fe/H]
$\pm$0.25). This is done for four stars covering a wide range
in atmospheric parameters.

\begin{table}
\caption[]{Dependence of abundances on stellar parameters}
\label{T1}
\begin{tabular}{rrrrrrrrrrrrr}
\hline \hline \noalign{\smallskip}
\multicolumn{1}{c}{}&\multicolumn{6}{c}{HD51530}&\multicolumn{6}{c}{HD15596}\\
\noalign{\smallskip} \hline \noalign{\smallskip}
\multicolumn{1}{c}{}&\multicolumn{6}{c}{(6100/3.8/0.8/--0.39)}
&\multicolumn{6}{c}{(4750/2.5/1.0/--0.67)}\\
\noalign{\smallskip} \hline
El/D& 1& 2& 3& 4& 5& 6& 1& 2& 3& 4& 5& 6\\
\hline
Cu &0.09&0.00&0.04&0.00&0.05&0.11& 0.07&0.04&0.14&0.04&0.05&0.17\\
Zn &0.06&0.01&0.06&0.00&0.07&0.11& 0.04&0.10&0.10&0.06&0.06&0.17\\
\hline \hline \noalign{\smallskip} \multicolumn{1}{c}{}&
\multicolumn{6}{c}{HD219617}&\multicolumn{6}{c}{HD122563}\\
\noalign{\smallskip} \hline \noalign{\smallskip}
\multicolumn{1}{c}{}& \multicolumn{6}{c}{(5870/4.0/1.5/--1.43)}&
\multicolumn{6}{c}{(4570/1.1/1.2/--2.5)}\\
\noalign{\smallskip} \hline
El/D& 1& 2& 3& 4& 5& 6& 1& 2& 3& 4& 5& 6\\
\hline
Cu & 0.07&0.00&0.00&0.00&0.05&0.09&  0.23&0.07&0.01&0.01&0.05&0.25\\
Zn & 0.03&0.04&0.01&0.00&0.09&0.10&  0.12&0.13&0.01&0.00&0.09&0.20\\
\hline
\hline
\end{tabular}
\begin{list}{}{}
\item[(1)] -- $\Delta$T$_{eff}$ = +100 K;
\item[(2)] -- $\Delta$$log$g= +0.3;
\item[(3)] -- $\Delta$V$_{t}$=+0.2 km s$^{-1}$;
\item[(4)] -- $\Delta$[Fe/H] = --0.25 dex;
\item[(5)] -- $\Delta$EW= $\pm$ 2m\AA ~for Zn and 0.05 dex for Cu
abundance;
\item[(6)] -- Total error.
\end{list}
\end{table}

Besides errors caused by model parameters, the total error
includes other uncertainties: mainly in the measurement of EWs
for zinc ($\pm$ 2m\AA) and mainly in the spectrum fit for copper
($\pm$ 0.05dex in Cu abundance). From the analysis of Table~2
we deduce that Cu is more sensitive to the choice of
$T_{eff}$ than Zn, while Zn is more sensitive to $\log$ g. Such a
behavior of zinc is due to its higher ionization potential. The
main contribution to the total error comes from the uncertainty in
$V_{t}$ determination for lines having EWs larger than 100 m\AA.
The total error is around 0.09-0.11dex for dwarf stars with high
T$_{eff}$ and grows when T$_{eff}$ and $\log$ g decrease,
reaching values of 0.20--0.25 dex for cool giants. On this basis
we estimate that the {\it average} accuracy of Cu and Zn abundances
is about 0.15 dex. Individual uncertainties for our sample stars
are listed in Table~3, together with the ratios of Cu and Zn
abundances relative to Fe.

%
\begin{table*}
\caption[]{Elemental abundances for the stars in our sample}
\begin{tabular}{lrrrrrrrrrrr}
\hline HD&[Fe/H]&[Cu/Fe]&$\sigma$
&[Zn/Fe]&$\sigma$&HD/BD&[Fe/H]&[Cu/Fe]&$\sigma$
&[Zn/Fe]&$\sigma$ \\
\hline
 245  & --0.78&  0.05&0.16&  0.15&0.13&117876  &--0.47&  0.17&0.15&
0.12&0.14\\
 2796 & --2.21&--0.54&0.14&  0.12&0.12&122563  &--2.66&--0.72&0.25&
0.12&0.19\\
 3546 & --0.63&--0.10&0.15&  0.22&0.14&122956  &--1.60&--0.75&0.16&
0.01&0.17\\
 3567 & --1.20&--0.66&0.13&--0.23&0.16&124897  &--0.58&  0.06&0.17&
0.25&0.15\\
 4306 & --2.52&--0.54&0.13&  0.04&0.12&127243  &--0.66&--0.07&0.21&
0.13&0.18\\
 5395 & --0.19&  0.09&0.19&  0.12&0.21&132142  &--0.54&  0.09&0.24&
0.16&0.22\\
 5916 & --0.51&    --&  --&  0.33&0.26&134169  &--0.72&
0.08&0.13&--0.02&0.11\\
 6582 & --0.89&  0.22&0.14&  0.22&0.15&140283  &--2.41&   -- &  --&
0.14&0.12\\
 6755 & --1.46&--0.79&0.19&--0.13&0.16&150177  &--0.65&--0.15&0.12&
0.01&0.11\\
 6833 & --1.13&--0.12&0.24&--0.14&0.26&157089  &--0.58&--0.02&0.14&
0.08&0.13\\
 8724 & --1.70&--0.60&0.22&--0.09&0.19&159482  &--0.86&--0.11&0.16&
0.19&0.14\\
 10700& --0.56&  0.21&0.19&  0.26&0.18&160693  &--0.46&  0.06&0.16&
0.04&0.14\\
 13530& --0.48&  0.00&0.15&  0.14&0.14&165195  &--2.03&--0.57&0.26&
0.05&0.22\\
 13783& --0.61&  0.14&0.17&  0.15&0.19&165908  &--0.67&  0.02&0.12&
0.06&0.11\\
 15596& --0.67&  0.22&0.28&  0.28&0.21&166161  &--1.20&--0.30&0.16&
0.11&0.13\\
 18768& --0.51&--0.12&0.25&  0.21&0.23&175305
&--1.47&--0.30&0.17&--0.08&0.14\\
 19445& --1.89&--0.54&0.10&--0.04&0.08&184499  &--0.66&  0.13&0.13&
0.10&0.12\\
 23439& --1.14&  --  &  --&   -- &  --&187111  &--1.74&--0.58&0.24&
0.14&0.20\\
 25329& --1.73&--0.37&0.11&  0.01&0.14&188510
&--1.48&--0.52&0.13&--0.03&0.12\\
 26297& --1.91&--0.72&0.32&  0.23&0.24&189558  &--1.00&--0.30&0.11&
0.09&0.11\\
 37828& --1.49&--0.41&0.25&  0.04&0.26&194598
&--1.16&--0.34&0.11&--0.10&0.10\\
 44007& --1.49&--0.61&0.17&--0.01&0.16&195633
&--0.55&--0.05&0.12&--0.05&0.11\\
 45282& --1.31&--0.48&0.13&--0.18&0.13&201889
&--0.78&--0.14&0.12&--0.06&0.13\\
 46480& --0.49&    --&  --&  0.31&0.21&201891
&--0.99&--0.21&0.11&--0.10&0.10\\
 51530& --0.39&--0.01&0.14&  0.13&0.13&204155  &--0.82&  0.05&0.15&
0.26&0.13\\
 63791& --1.73&--0.52&0.28&  0.17&0.20&204543  &--1.79&--0.60&0.28&
0.07&0.20\\
 64090& --1.69&--0.51&0.16&--0.01&0.17&208906
&--0.71&--0.09&0.11&--0.10&0.10\\
 64606& --0.99&  0.09&0.10&  0.21&0.11&216143  &--2.13&--0.57&0.32&
0.08&0.24\\
 76932& --0.90&--0.17&0.15&  0.21&0.14&216174  &--0.51&
0.18&0.26&--0.11&0.26\\
 84937& --2.00&   -- &  --&  0.09&0.11&218502  &--1.72&   -- & -- &
 -- & -- \\
 87140& --1.71&--0.54&0.14&  0.09&0.15&218857
&--1.84&--0.76&0.24&--0.05&0.20\\
 88609& --2.66&--0.64&0.25&  0.16&0.19&219617  &--1.43&   -- & --
&--0.04&0.09\\
 88725& --0.71&  0.09&0.13&  0.19&0.15&221170  &--2.05&--0.70&0.24&
0.05&0.20\\
 94028& --1.43&--0.47&0.11&--0.03&0.09&221377
&--0.90&--0.20&0.11&--0.12&0.10\\
103095& --1.47&--0.40&0.14&  0.15&0.16&224930
&--0.85&--0.07&0.13&
0.33&0.14\\
105755& --0.65&--0.14&0.13&  0.07&0.09&338529  &--2.31&   -- & --
&
0.15&0.25\\
108076& --0.85&--0.02&0.17&  0.25&0.14&345957
&--1.33&--0.27&0.13&
0.00&0.16\\
108317& --2.17&--0.63&0.13&  0.01&0.11&-18 5550&--3.01&   -- & --
&--0.03&0.26\\
110184& --2.27&--0.63&0.31&  0.22&0.24&+02 3375&--2.26&   -- & --
&
 -- & -- \\
114762& --0.74&--0.06&0.13&  0.09&0.11&+02 4651&--1.82&   -- & --
&
0.13&0.19\\
\hline
\end{tabular}
\end{table*}

%
\begin{table*}
{Table 3 (Continued)}\\
\begin{tabular}{lrrrrrrrrrrr}
\hline BD&[Fe/H]&[Cu/Fe]&$\sigma$
&[Zn/Fe]&$\sigma$&BD&[Fe/H]&[Cu/Fe]&$\sigma$
&[Zn/Fe]&$\sigma$ \\
\hline +04 4551&--1.51&   -- & -- &  0.02&0.14&+30
2611&--1.41&--0.75&0.16&--0.06&0.13\\
+17 4708&--1.56&   -- & -- &--0.05&0.16&+36 2165&--1.51&   -- &
-- &
 0.01&0.26\\
+23 3130&--2.61&   -- & -- &  0.11&0.10&+41
3931&--1.68&--0.97&0.22&--0.06&0.19\\
+29 0366&--1.01&--0.14&0.26&  0.25&0.24&+42 3607&--1.97&   -- & --
&--0.04&0.11\\
+29 2091&--1.93&   -- & -- &  0.29&0.20&+66 0268&--1.95&   -- &
-- &
 0.11&0.22\\
\hline
\end{tabular}
\end{table*}

Fig.~2 to 5 show a Cu and Zn abundances as a function of T$_{eff}$,
$\log$ g. A general lack of correlation emerges, and this confirms
that the choice of parameters is reliable. We also compared our
[Cu/Fe] and [Zn/Fe] ratios with those by Sneden et al. (1991). This
is done in Fig.~6 and 7. The mean difference between our results
and those of Sneden et al. (\cite{sgc91}) for the stars in common
is $\Delta$[Cu/Fe]=--0.02 $\pm$0.14 (15 stars) and
$\Delta$[Zn/Fe]=0.01$\pm$0.10 (23 stars), respectively. The
agreement is therefore quite good.

Finally, a few comments on nLTE effects are needed. These
effects in spectral lines are caused by both the real conditions
in the star's atmosphere and the structure of atomic levels.
Non-LTE corrections require detailed calculations for each case
considered. Building an adequate atomic model is a
difficult task, especially in the case of complex structures of
atomic terms and unreliable atomic parameters
(in presence of photoionization, shock processes etc.). Concerning Cu and Zn,
computations of nLTE-effects on their abundances are complex and
so far a good estimate of nLTE-corrections has not been
presented. A posteriori, obtaining different abundances from lines
with different low-level excitation potentials can be an empirical
demonstration that departures from LTE are present.  In the case
of Cu, the lines 5105 \AA~ and 5782 \AA~ occur from meta-stable
levels (E$_{low}$ = 1.39 and 1.64 eV, respectively), unlike the
line at 5218 \AA (E$_{low}$ = 3.82 eV). We have estimated that
the discrepancy in Cu abundance, as determined from the lines at
5105 \AA~ and at 5218 \AA, is $<$ [Cu/H]$_{5105}$ -
[Cu/H]$_{5218}$ $< =$ 0.04$\pm$0.10. The difference is well
within the uncertainty of the estimate, and does not provide
reasons for assuming that departures from LTE are
of any importance. Unfortunately, Zn I lines have similar potentials
and we cannot perform a similar analysis for the Zn abundance. We can
however note that we used solar oscillator strengths for Zn I
lines  from Gurtovenko \& Kostyk~(1989), which implicitly include
departures from LTE. We thus consider that our results can be
affected only marginally by nLTE- effects.

%
\begin{figure}
\vspace{1cm} \resizebox{\hsize}{!}{\includegraphics{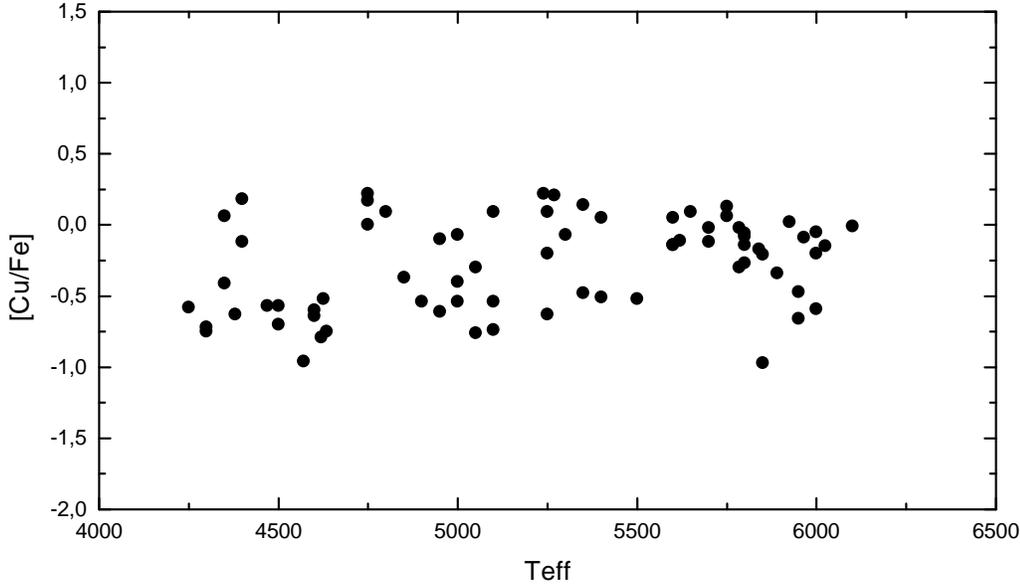}}
\caption[]{[Cu/Fe] abundance ratios as a function of $T_{eff}$
for our sample stars.} \label{Fig2}
\end{figure}
%

\begin{figure}
\vspace{1cm} \resizebox{\hsize}{!}{\includegraphics{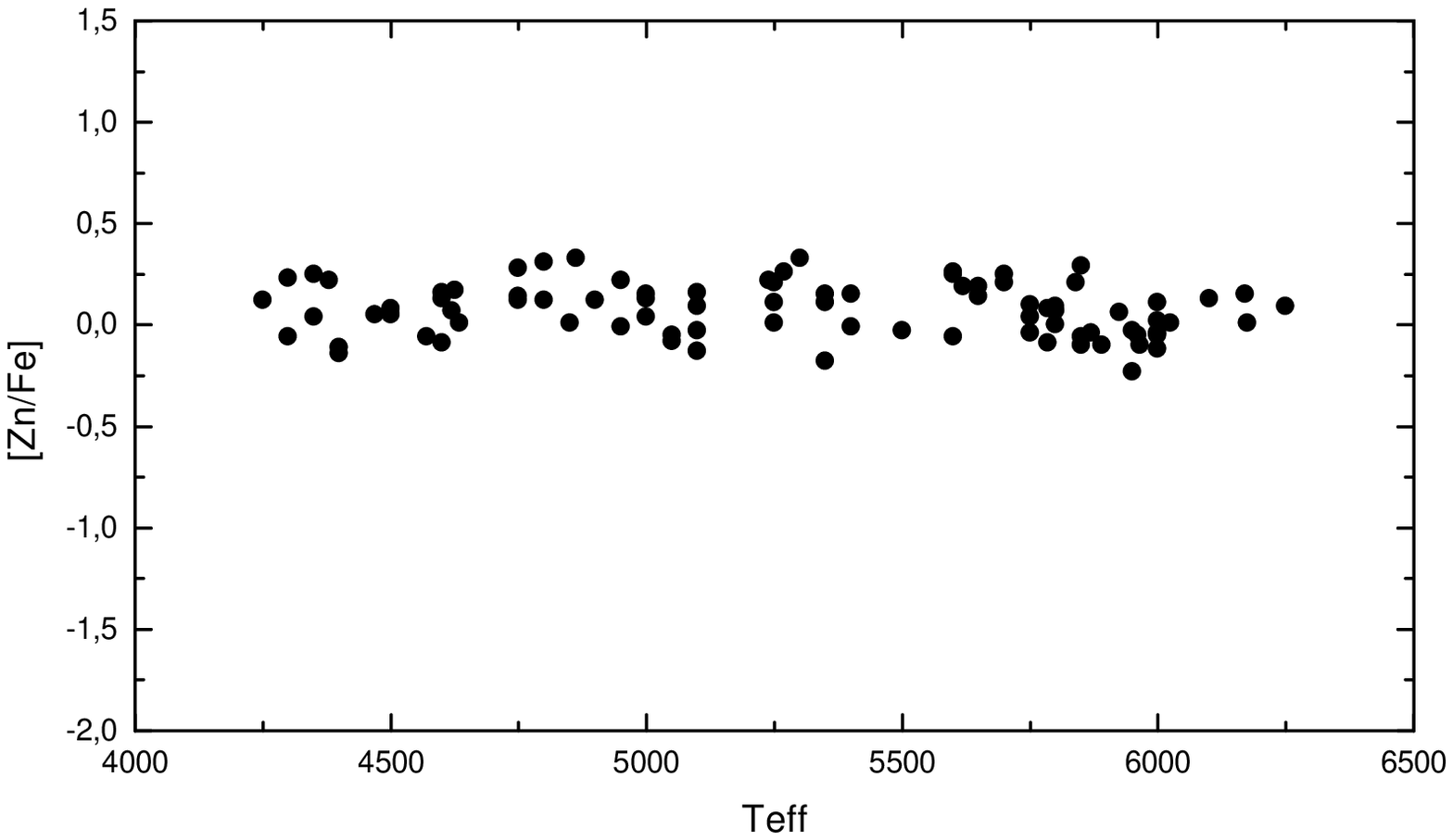}}
\caption[]{[Zn/Fe] abundance ratios as a function of $T_{eff}$ for
our sample stars.} \label{Fig3}
\end{figure}
%

%
\begin{figure}
\vspace{1cm} \resizebox{\hsize}{!}{\includegraphics{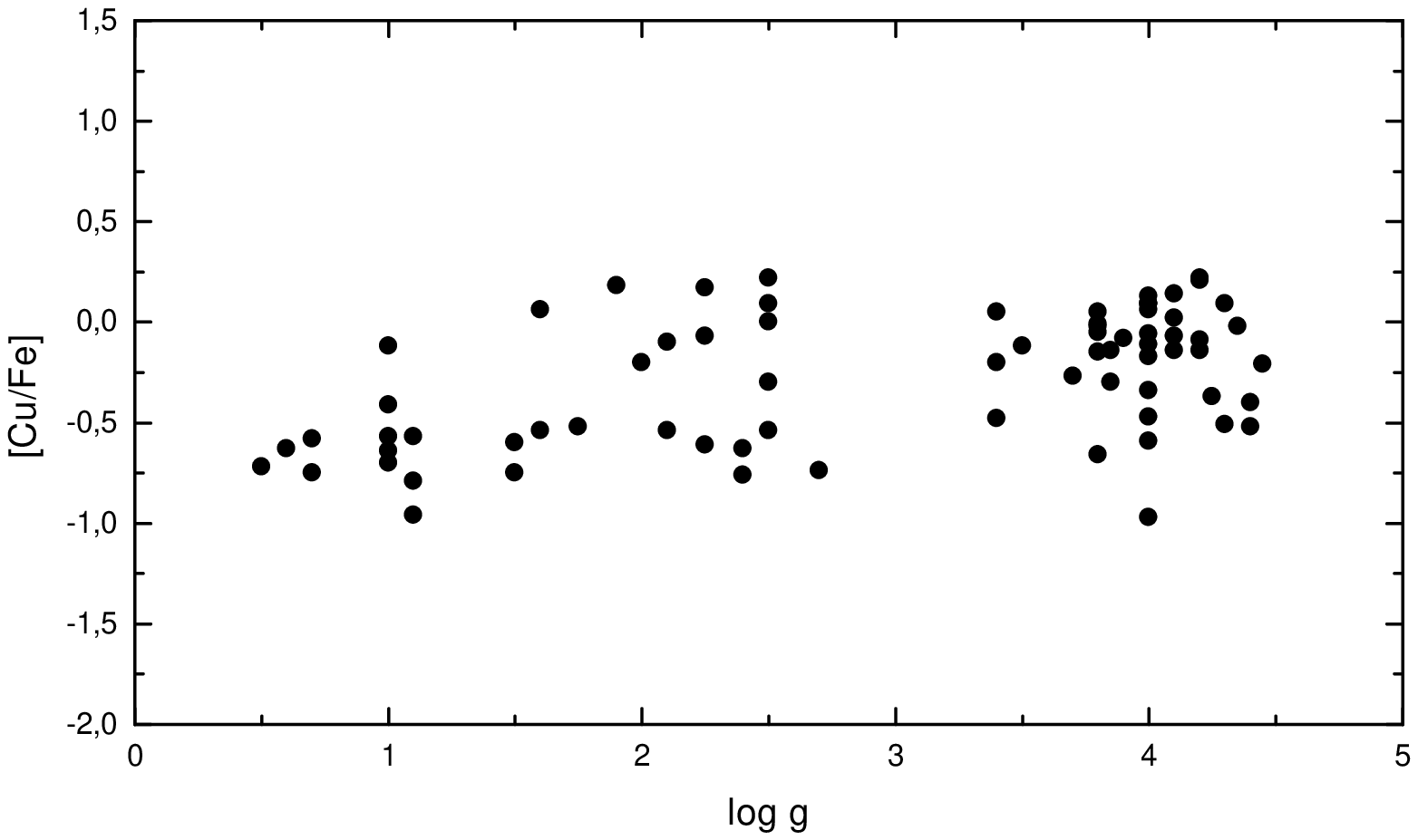}}
\caption[]{[Cu/Fe] abundance ratios as a function of $\log$ g for
our sample stars.} \label{Fig4}
\end{figure}
%

\begin{figure}
\vspace{1cm} \resizebox{\hsize}{!}{\includegraphics{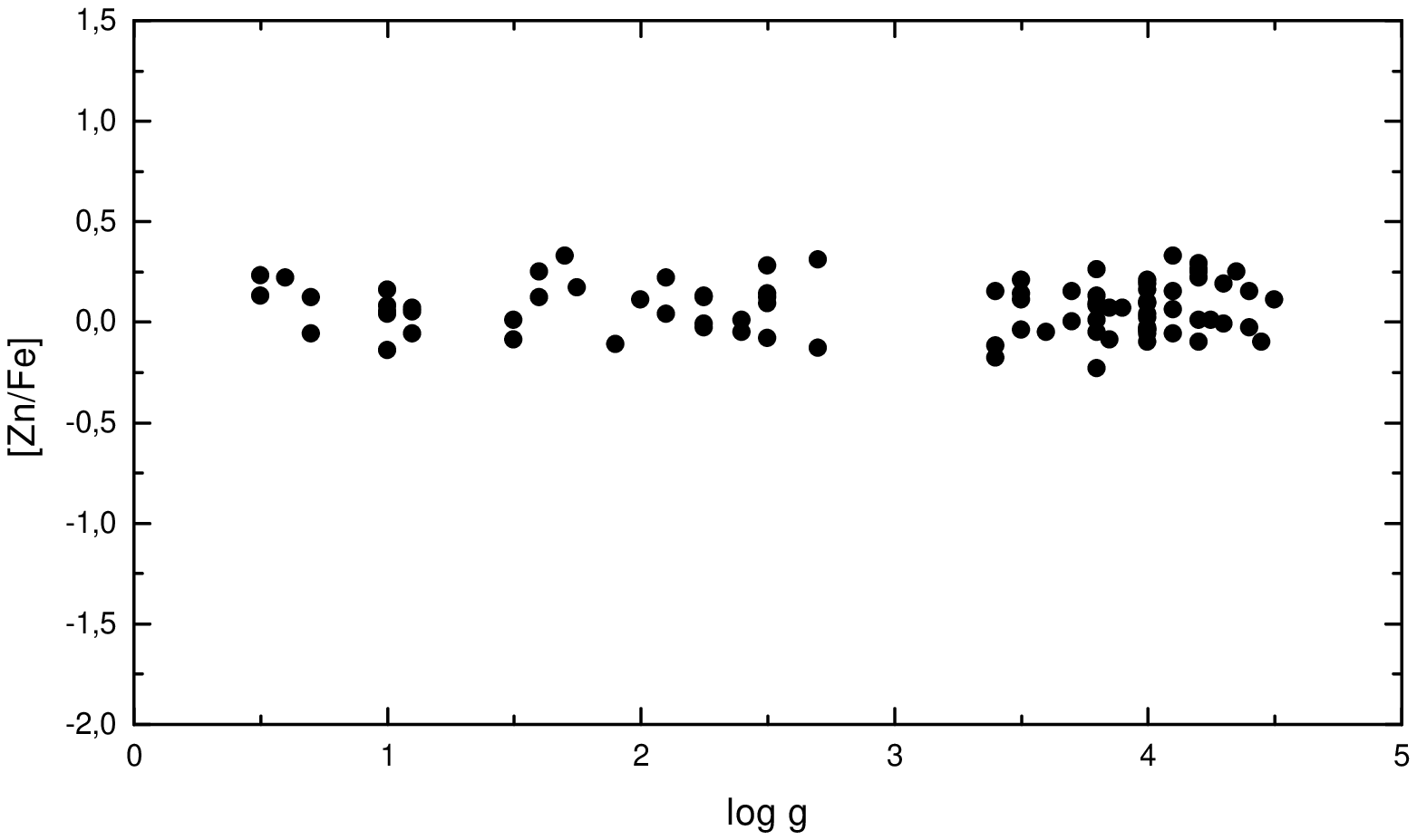}}
\caption[]{[Zn/Fe] abundance ratios as a function of $\log$ g for
our sample stars.} \label{Fig5}
\end{figure}
%

%
\begin{figure}
\vspace{1cm} \resizebox{\hsize}{!}{\includegraphics{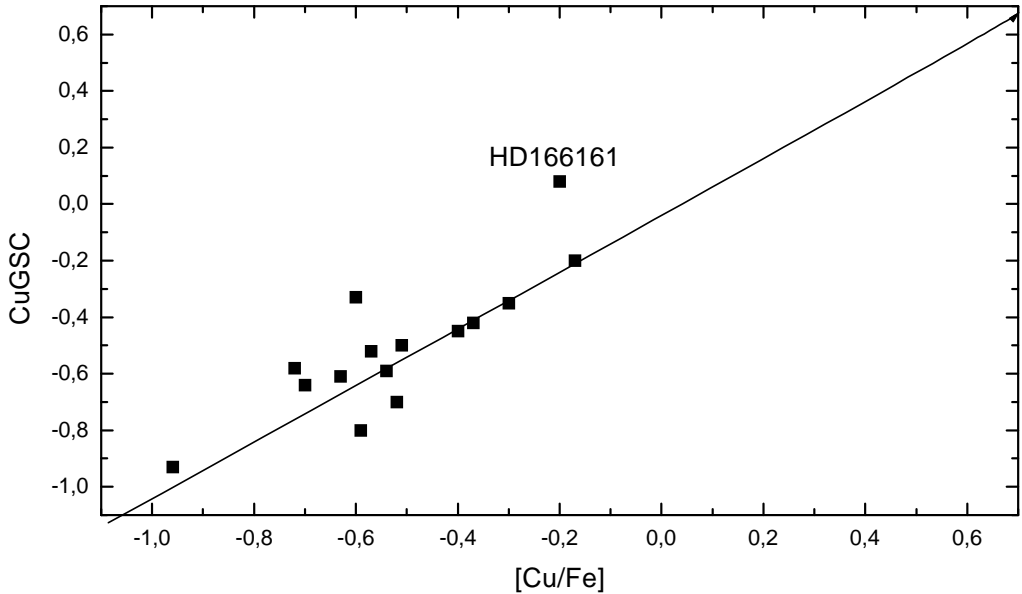}}
\caption[]{A comparison of our [Cu/Fe] abundance ratios with
those by Sneden et al.~(1991). Some discrepancy for HD 166161 is
due to the different model parameters used in the two works.
Uncertainties for this star are known to be due to its large
reddening (Sneden et al.~1991).} \label{Fig6}
\end{figure}
%

\begin{figure}
\vspace{1cm} \resizebox{\hsize}{!}{\includegraphics{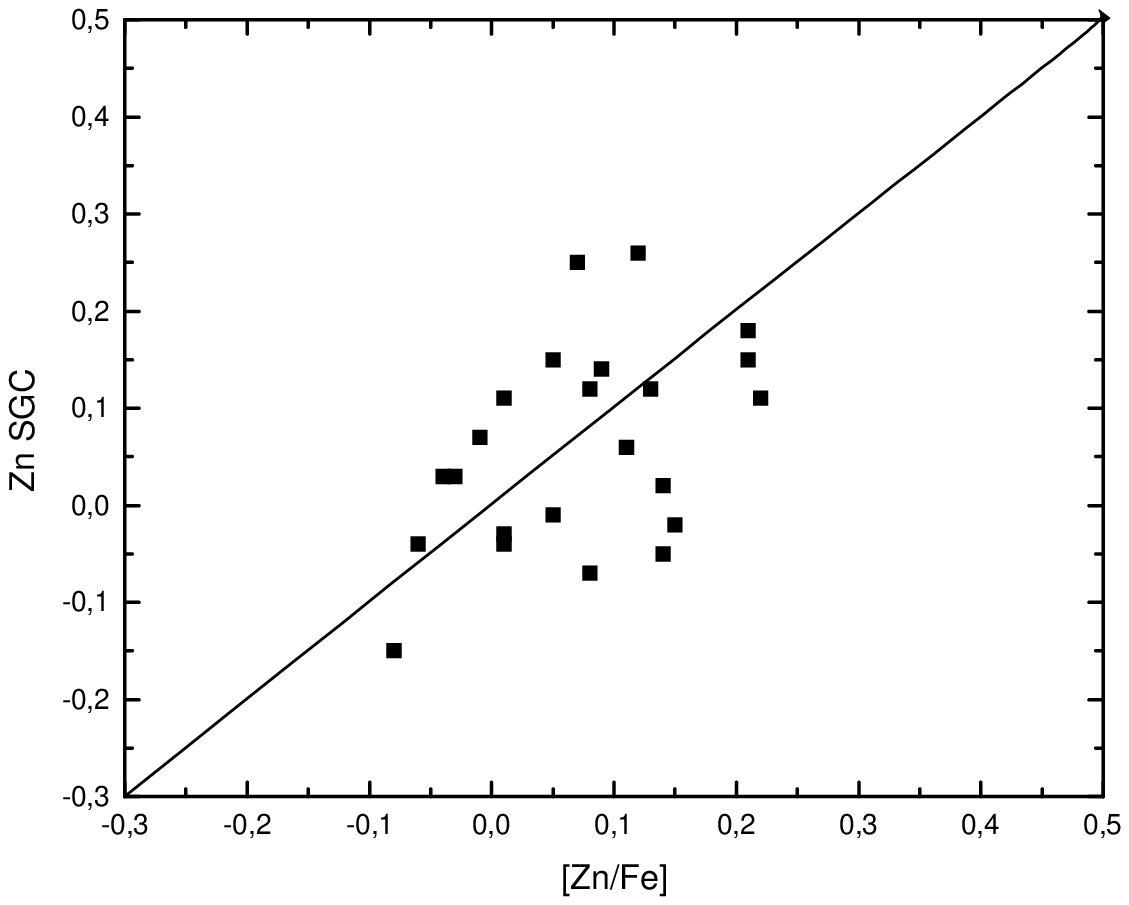}}
\caption[]{A comparison of our [Zn/Fe] abundance ratios with
those by Sneden et al.~(1991)} \label{Fig7}
\end{figure}

\subsection{Resulting abundance trends}

Our values of [Cu/Fe], [Zn/Fe], and [Cu/Zn] are shown as a
function of [Fe/H] in Fig.~8, and compared to abundances available
in the literature. In the figure we prefer to use different points for
measurements of the same star made by different authors, instead
of making averages: in this way the scatter related to the
heterogeneous composition of the database can be better
appreciated. The trends for Cu and Zn are remarkably different,
in good accord with earlier findings by Gratton \& Sneden (1988)
and by Sneden et al. (1991), suggesting that differences in the
dominant stellar mechanisms controlling the production of these
elements must exist.

\begin{figure}
\vspace{1cm} \resizebox{\hsize}{!}{\includegraphics{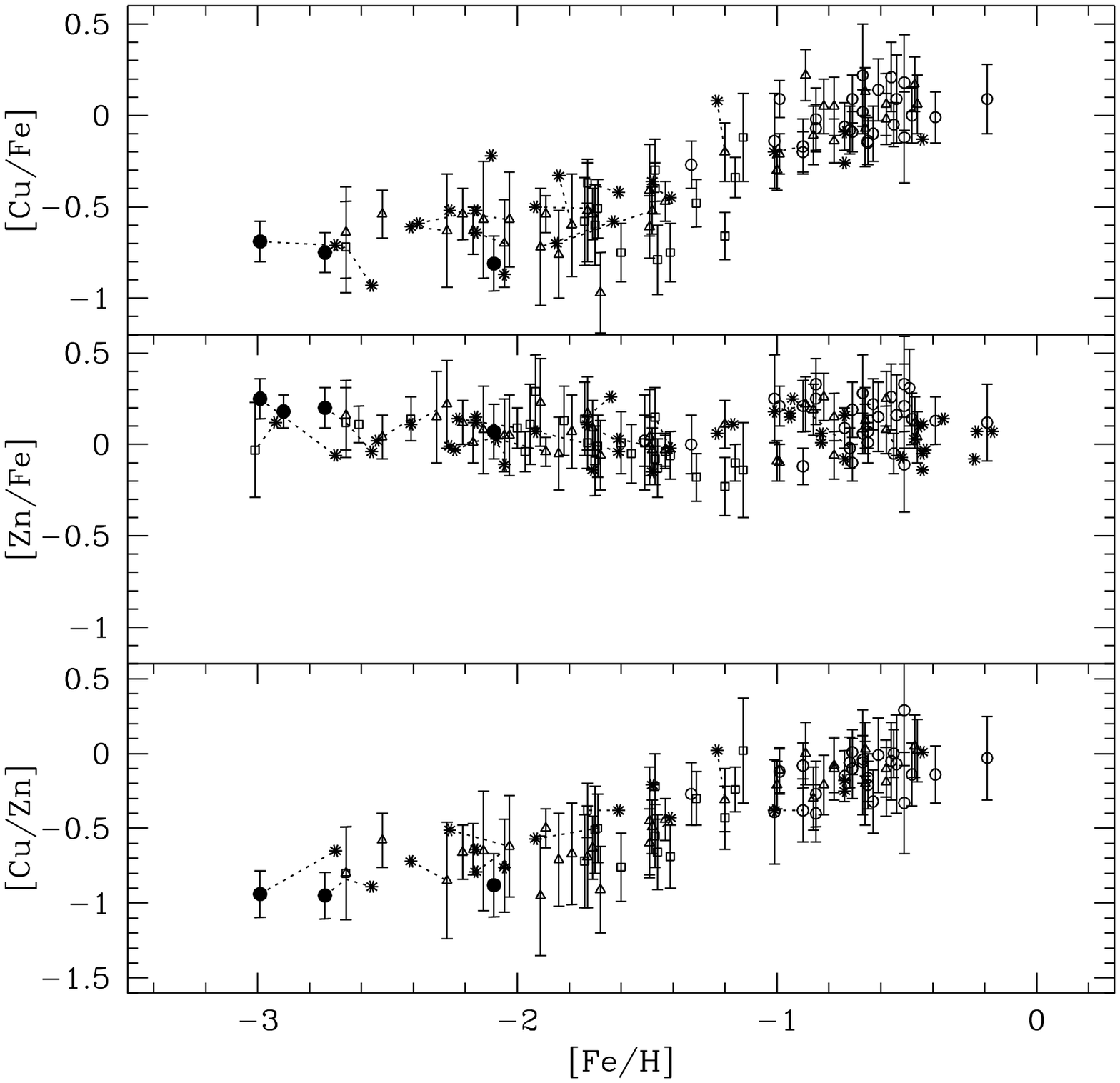}}
\caption[]{Observational trends of [Cu/Fe] ({\it upper panel}),
[Zn/Fe] ({\it middle panel}), and [Cu/Zn] ({\it lower panel})
versus [Fe/H] are shown. Symbols are for: H stars ({\it open
triangles}), D ({\it open circles}), and I stars ({\it open
squares}) from this work; Sneden et al.~(1991) ({\it stars});
Westin et al.~(2000), Hill et al.~(2002), and Cowan et al.~(2002)
({\it filled circles}). Thin dotted lines connect the points
representing observations of the same stars by different authors.
Error bars for individual objects are shown only when reported
in the original papers.} \label{Fig8}
\end{figure}

The mean values of [Cu/Fe] and [Zn/Fe] for thick-disk (D) and halo
(H) stars are --0.10$\pm$0.23 (29 stars, D)  and --0.58$\pm$0.25
(15 stars, H); 0.14$\pm$0.14 (34 stars, D) and --0.03$\pm$0.11 (23
stars, H), respectively.  The abundance of Cu is therefore
remarkably lower for halo stars than for disk stars, and this
requires contributions on relatively long time scales. We shall
comment on this point in section 4. On the contrary, Zn abundances
have a flat trend, with a possible indication of a slight
overabundance in the halo (at the limit of the error bar): this
in general confirms the well-known fact that the ratio [Zn/Fe] is
almost solar at all metallicities.

\section{Nucleosynthesis and chemical evolution of Cu and Zn}

Let us discuss now what the Cu and Zn abundance measurements in
metal-poor stars imply for their nucleosynthesis and chemical
evolution. As mentioned, this topic was previously addressed by Timmes
et al.~(1995) and by Matteucci et al.~(1993). The first authors
adopted nucleosynthesis predictions by Woosley \& Weaver
(\cite{ww95}). According to their calculations, the dominant
contribution to copper derives from Ne-burning in massive stars
exploding as Type II supernovae, while Zn is formed mainly during
explosive Si-burning. When our enlarged database is considered,
Timmes et al.~(1995)'s model appears no longer compatible with
the observations and must now be ruled out. The
models by Matteucci et al.~(1993), including a relevant
contribution from Type Ia supernovae, yield predictions
somewhat closer to the data, but are nevertheless not
sufficient, in particular for Cu at low metallicity.

Despite the many studies on stellar nucleosynthesis published
in the past years, the scenario for the production
of Cu and Zn remains unclear. Actually, not only the situation
has not improved significantly, but we are now aware that
this failure is related to some fundamental weakness
in the computations of pre-SN and explosive nucleosynthesis
of nuclei after the iron peak. Indeed, a remarkable overproduction
of such nuclei seems to be at present unavoidable. Revisions in
the input physics and in reaction rates have somewhat alleviated
the difficulties, as compared to the original calculations by
Woosley \& Weaver~(1995), but the problems are not solved
(Hoffman et al.~2001; Heger et al.~2001; Rauscher et al.~2002).
Therefore, predictions for the synthesis of Cu and Zn in stars
ultimately evolving to Type II supernovae must be viewed with
great care. A similar (or even worst) situation occurs for Type Ia
supernovae (see e.g.Nomoto, Thielemann, \& Wheeler~1984; Iwamoto
et al.~1999; Brachwitz et al.~2000), and the only contributions
relatively well established are those from the main component of the
$s$-process coming from long-lived AGB stars (Gallino et
al.~1998; Busso et al.~2001). However, these contributions to Cu
and Zn are marginal.

\subsection{Production of Cu and Zn: hints from observations}

Starting from Cu, we have mentioned that the trend of Fig.~8 can
be interpreted accepting that Cu owes a larger fraction
of its abundance to long-lived mechanisms than iron does.
The data actually say something more, i.e. that the two
elements are not linked by a simple relation. Fig.~9 shows indeed
that fitting their relative trends requires at least a polynomial
function with a quartic term. The line in the figure has no
special meaning other than minimizing the sigma of the fit; it
however shows that any reasonably accurate interpolation curve
must necessarily assume relationships more complex than expected
from purely primary or purely secondary processes.

\begin{figure}
\vspace{1cm} \resizebox{\hsize}{!}{\includegraphics{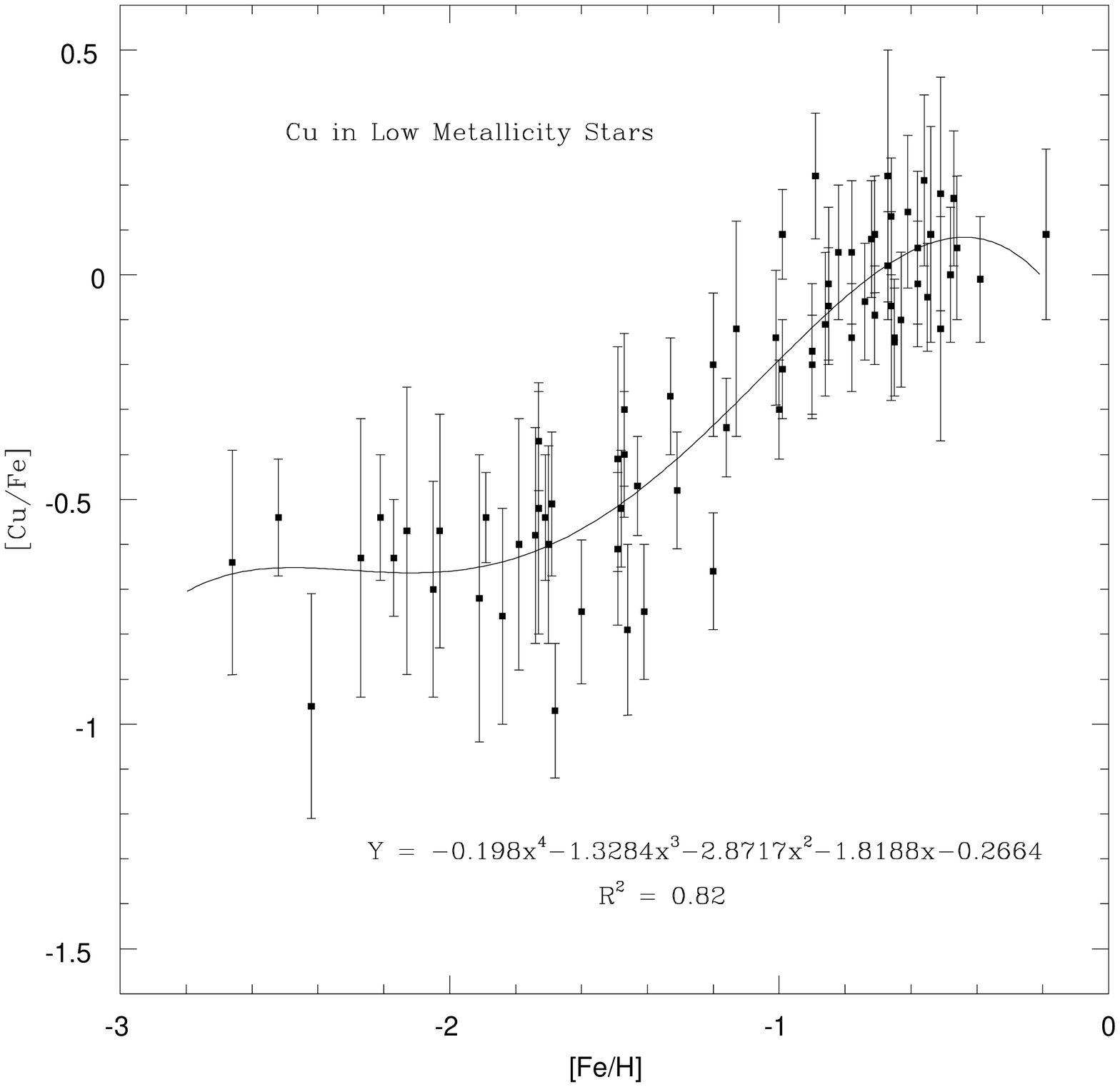}}
\caption[]{[Cu/Fe] vs. [Fe/H]: {\it filled squares} are the
observations of Fig.~8. The fit ({\it continuous line}) shows that
the dependency of [Cu/Fe] on [Fe/H] is more complex than
implied by purely primary or purely secondary mechanisms
(see text for discussion).}
\label{Fig9} \end{figure}
%
\begin{figure}
\vspace{1cm} \resizebox{\hsize}{!}{\includegraphics{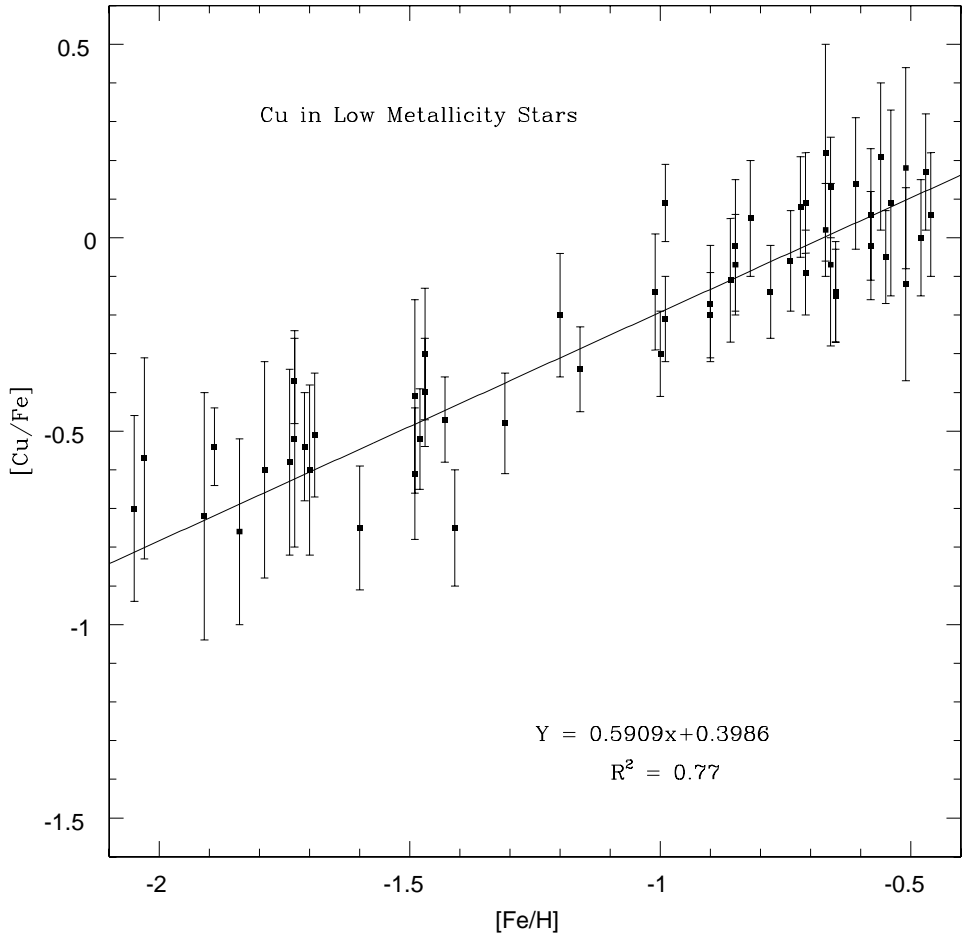}}
\caption[]{[Cu/Fe] abundances in the range  $-$2 $\le$ [Fe/H]
$\le$ $-$0.5: observations ({\it filled squares} are the same as
in Fig.~8). The linear relation ({\it continuous line}), fitting
[Cu/Fe] in this metallicity range, has a slope close to 0.6, and
is incompatible with a simply secondary behavior.}
\label{Fig10}
\end{figure}
%
\begin{figure}
\vspace{1cm} \resizebox{\hsize}{!}{\includegraphics{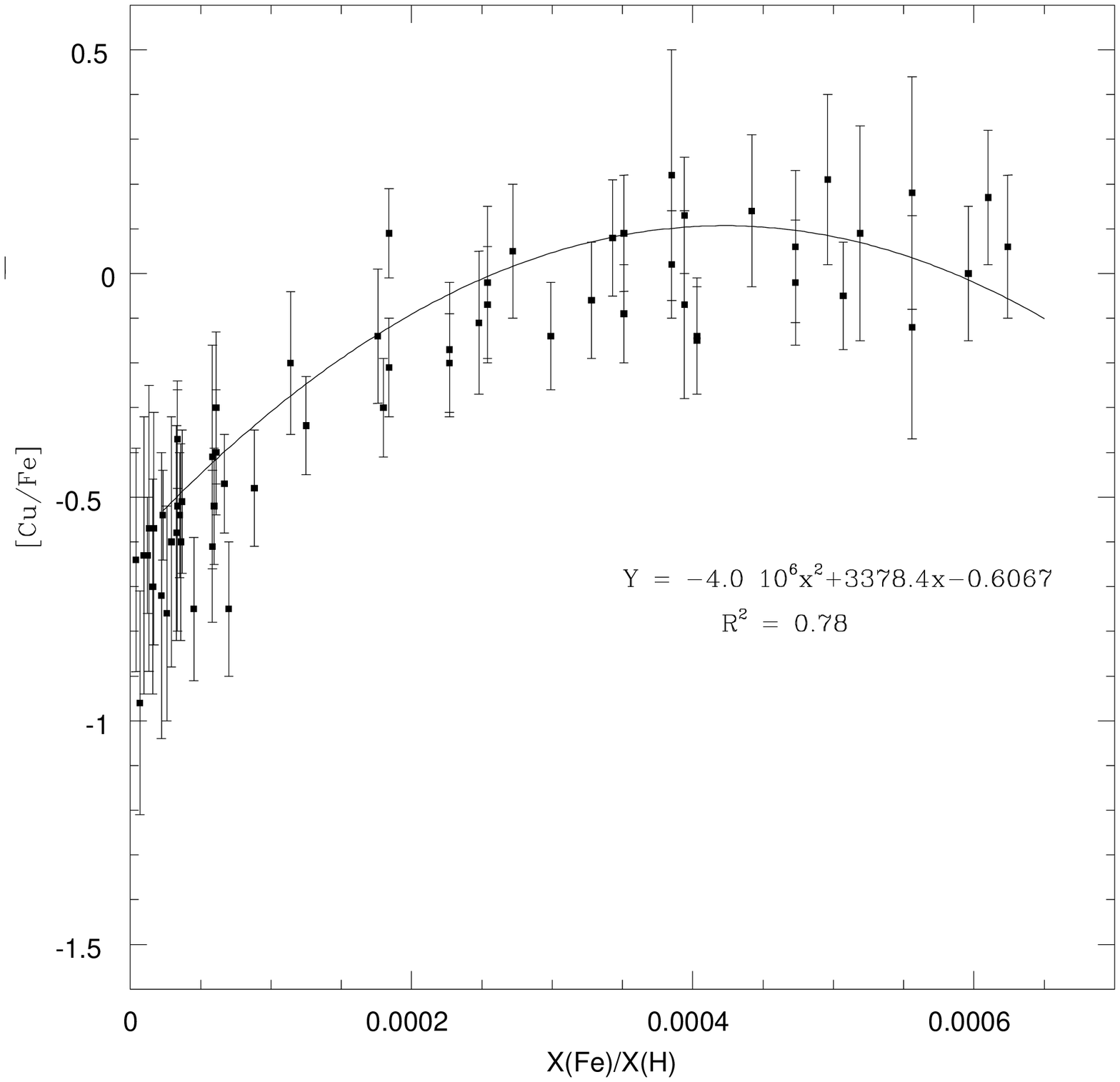}}
\caption[]{[Cu/Fe] abundances ({\it filled squares}, showing the
same data as in Fig.~8), plotted versus the linear ratio
X(Fe)/X(H). The trend is parabolic, and might result from a
superposition of primary and secondary components.}
\label{Fig11}
\end{figure}
In order to show this and infer what kind of phenomena might be
involved, let us remember that, when the stellar lifetime is
short enough that it can be neglected in calculations of galactic
chemical evolution, a secondary element $S$ reaches an abundance:
$$
X^S = 1/2 y^S y^{Fe} [f(t)]^2 \eqno(1)
$$
where y$^S$ is the so-called {\it integrated effective yield} of
element $-S$, produced by a whole generation of stars. If we
indicate by $\mu$ the residual galactic mass in the form of gas,
f(t) = [ln$\mu^{-1}$] for closed-box models (galactic halo) and
f(t) = [ln$\mu^{-1}$-1] for infall models (galactic disk: see e.g.
Lynden-Bell~1975, Chiosi~1986, for more extended definitions and
treatments). A primary element like Fe (or, better, its
contribution by short-lived objects whose lifetime can be
neglected) reaches instead an abundance:
$$
X^{P} = y^{P} [f(t)] \eqno(2)
$$
so that the ratio $X^S/X^{Fe}$ is proportional to f(t), hence to
the ratio $X^{Fe}/y^{Fe}$. Computing logarithms, it is
straightforward to recognize that [S/Fe] =
[Fe/H] + const.: hence in Fig.~9 a purely secondary element
would have a linear trend with slope +1. On the contrary, in the
region where an almost linear trend exists ($-$2 $\le$ [Fe/H] $\le$
$-$0.5) the observed slope is about 0.6 (see also Fig.~10). Hence,
already in early times of galactic evolution (the halo phases) Cu
cannot be considered as a purely secondary element. In such phases
it might be explained as the superposition of (at least) two independent
processes, one of primary and one of secondary origin. Indeed,
summing two components defined by equations (1) and (2), one gets:
$$
[Cu/Fe] = A + B\times(X(Fe)/X(H)) + C\times(X(Fe)/X(H))^2 \eqno(3)
$$
where A, B, C are known functions of the solar Cu and Fe abundances,
and of the effective integrated yields of Fe, of Cu (primary), and of
Cu (secondary). A parabolic relation like the one in equation (3)
does in fact hold for the Cu abundance data in the shown metallicity
interval, as can be seen in Fig.~11. Hence, Cu can have contributions
from both primary and secondary processes in short-lived (i.e. massive)
stars. The importance of secondary mechanisms can be quantitatively
estimated by considering Cu production compared to the $s$-only
isotope $^{80}$Kr. This nucleus is known to be produced at about 12\%
of its abundance in the main component of the $s$-process
(Arlandini et al.~1999) and to have possibly a 10\% p-process
contribution. The rest is due to neutron captures in massive
stars. Let us use for these last the recent models by Hoffman et
al.~(2001), where the overproduction of isotopes up to A = 100 is
given. By weighting the predictions from the 15, 20 and 25
$M_{\odot}$ stars of the quoted paper over an initial mass
function $\Psi (M) \sim M^{-2}$, it is easy to see that {\it if}
those models have to produce $^{80}$Kr at the 78\% level of its
solar abundance, then summing over the isotopes of Cu one gets
for this element a contribution from secondary processes in
massive stars of about 23\% (assuming no primary contribution
from the same stars). Note that the predictions by Raiteri et
al.~(1992) and Matteucci et al.~(1993) are in almost perfect
agreement with this apparently weak argument (they found at most
26\%). This gives us some confidence in the result. Below [Fe/H]
= $-$2.5, [Cu/Fe] reaches an apparent plateau (though this
conclusion is uncertain due to the limited data) around $-$0.6
dex. As 70\% of iron comes later from type Ia Supernovae, this
implies a primary contribution by population II massive stars of
10$^{-0.6} \times$ 0.3, which gives a total contribution as small as
7.5\%. From the recent work by Heger et al.~(2002) we also see
that pre-galactic very massive stars cannot give remarkable
contributions to Cu.

We can also exclude that the primary massive star contribution be
dominated by the $r$-process. Inspection of Fig.~12, where the
trends of Cu and Zn versus Ba are shown, reveals indeed that at
low metallicity Cu and Zn remain high, maybe even growing with
respect to Ba. In these early phases of galactic evolution the
abundance of Ba is due to the $r$-process, which sharply
decreases below [Fe/H] = $-$2.5 (Travaglio et al.~1999, and
references therein). Cu and Zn behave differently and their
primary-like contribution from massive stars cannot be controlled
by the same mechanisms (and probably derives at least in part from
other explosive phenomena, like the $\alpha$-rich freeze-out of
NSE processes). We underline (though it should be obvious) that we
cannot exclude from such very crude arguments that the $r$-process
gives some contribution (see below). At this level we can only
state that it is not dominant.

\begin{figure}
\vspace{1cm} \resizebox{\hsize}{!}{\includegraphics{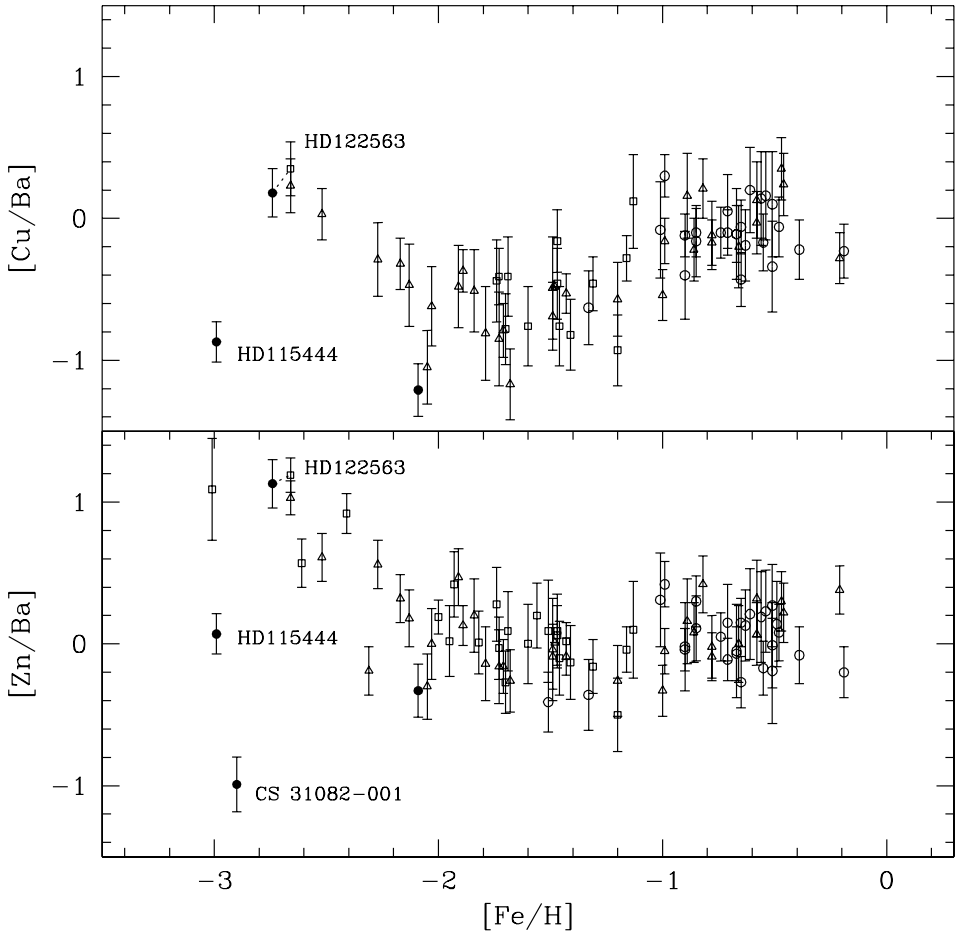}}
\caption[]{Trends of [Cu/Ba] ({\it upper panel}) and [Zn/Ba] ({\it
lower panel}), versus [Fe/H] are shown. Symbols are the same as in
Fig.~8. Again, thin dotted lines connect stars with different abundance
measurements, and individual error bars are shown only when reported
in the original works.} \label{Fig12}
\end{figure}
Summing up, inspection of the observed data, comparisons with
other elements and isotopes of known origin, and a very rough
scheme for the chemical enrichment of the galactic halo allows us
to predict that Cu receives a primary contribution by massive
stars of about 7.5\% of its abundance, while something around 25\%
should come from secondary processes in the same stars (slow
neutron captures, or the weak $s$-process). Another 5\% has been
already attributed to the main $s$-component from AGB stars of
long lifetime. We cannot avoid the suggestion that the remaining
part (formally 62.5\%) comes from the less known processes we
have so far neglected, i.e. explosive nucleosynthesis in Type Ia
supernovae. These would affect the abundance of Cu relatively
late, due to the longer lifetimes, and would thus contribute
to make the functional relationship between [Cu/Fe] and [Fe/H]
rather complex, as shown in Fig.~9. In an independent way, we
thus reach conclusions qualitatively similar to (though
quantitatively different from) Matteucci et al.~(1993).

A similar reasoning can be repeated for Zn, which however shows a
trend very close to Fe itself, so that the conclusion is more
straightforward. Attributing 3\% of its production to the main
$s$-process component in AGB stars, the rest should come either
from primary nucleosynthesis in massive stars ($\sim$30\%), or
from  Type Ia supernovae (something around 67\%, like for Fe). The
massive star yield cannot be dominated by the $r$-process, but
some contribution from this last remains possible.

\subsection{Galactic enrichment of Cu and Zn: a test and some
conclusions}

With the prescriptions derived in the previous section from simple
inspections of the data and comparisons with other nuclei of
known origin, we want now to compute the chemical evolution of
the Galaxy for Cu and Zn. This has obviously some circularity in
the procedure, and indeed we do not attribute any special emphasis
to such results: they are only an internal consistency check for
our predictions, when a full evolution model is adopted, based on
finite stellar lifetimes, and on global hypotheses for star
formation and nucleosynthesis of other elements that were already
tested in previous works.

For this 'test' we make use of a metallicity distribution and a
Star Formation Rate previously obtained (Travaglio et al.~1999)
through an evolutionary model suitable for reproducing a large
set of Galactic and extragalactic constraints (for details on the
code see Ferrini et al.~1992; for its application to galactic
heavy element enrichment see also Travaglio et al. 2001). The
model considers the Galaxy as divided in three zones, halo, thick
disk and thin disk. For Cu and Zn, we simply mimic the input
stellar yields by imposing the tentative production sites derived
in the previous section for the primary and secondary contributions
in massive stars, in Type Ia supernoave and in AGB stars.

Fig.~13 and Fig.~14 show the resulting chemical enrichment,
through plots of [Cu/Fe], [Zn/Fe], [Cu/Zn] and of [Cu/Ba], [Zn/Ba]
versus [Fe/H]. Chemical evolution predictions for Ba are from
Travaglio et al.~(1999). Being an outcome of the very crude
estimates previously discussed, the results illustrated by the
figures are not bad, and confirm that our suggestions for the
stellar yields should be roughly correct. However, while they can
interpret the general trends, they cannot explain the spread of
abundances at very low metallicity. This is in particular related
to the over-simplification of attributing massive star yields of
Cu and Zn to generic explosive phenomena, in the absence of a
criterion for distinguishing different processes occurring
in different masses.

\begin{figure}
\vspace{1cm} \resizebox{\hsize}{!}{\includegraphics{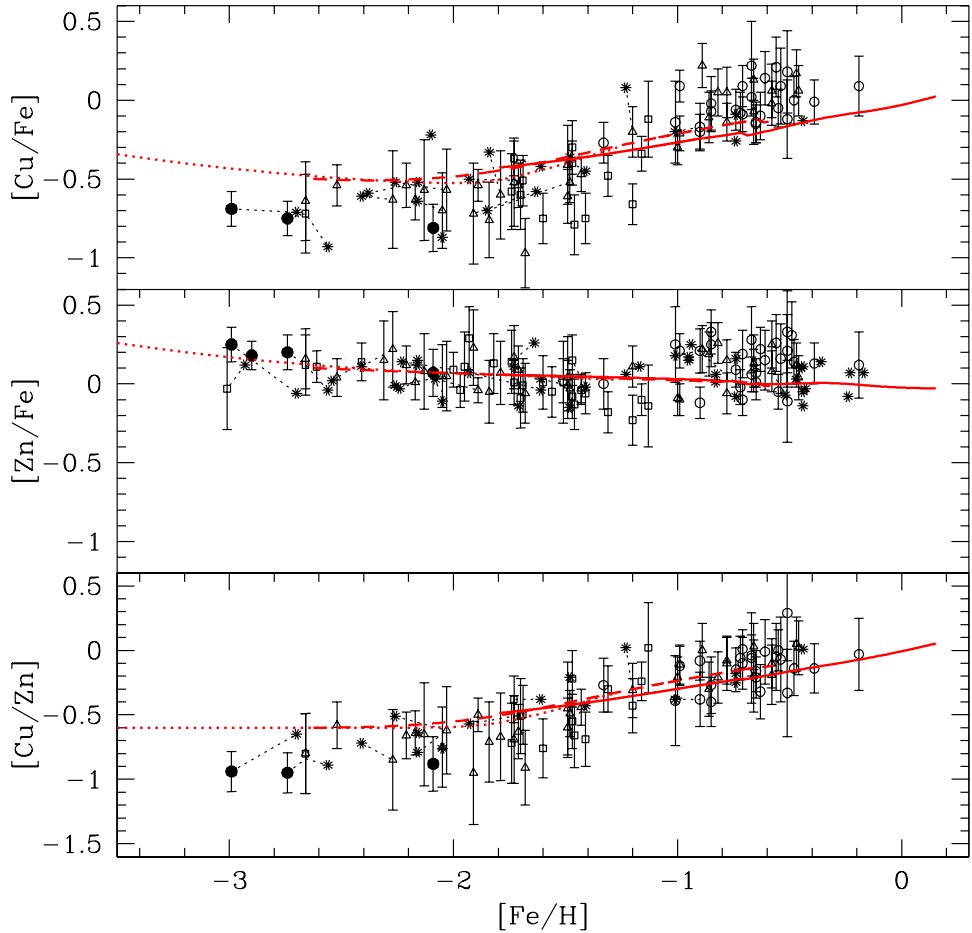}}
\caption[]{Galactic evolution of [Cu/Fe] ({\it upper panel}),
[Zn/Fe] ({\it middle panel}), [Cu/Zn] ({\it lower panel})
according to the chemical evolution prescriptions described in the
text (including primary processes from massive stars, secondary
processes from SNII, $s$-processes from AGB stars, and SNIa
contributions, in the relative fractions discussed in the text.
Symbols are the same as in Fig.~8.} \label{Fig13}
\end{figure}
%

\begin{figure}
\vspace{1cm} \resizebox{\hsize}{!}{\includegraphics{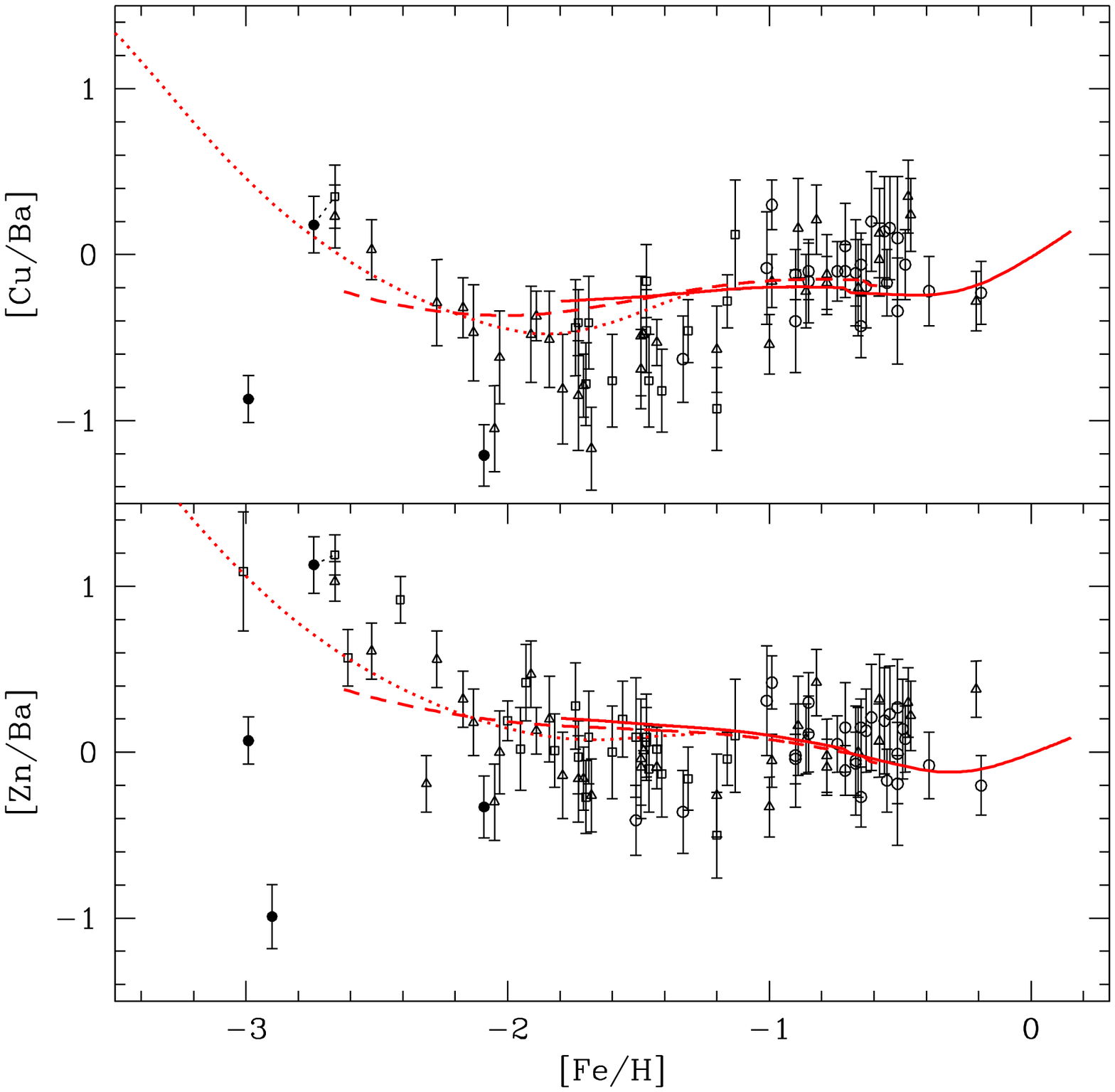}}
\caption[]{The same as Fig.~13 for the ratio [Cu/Ba] ({\it upper
panel}), and [Zn/Ba] ({\it lower panel}). Symbols are the same as
in Fig.~8.} \label{Fig14}
\end{figure}

In the galactic evolution results presented by Travaglio et
al.~(1999), the $r$-process was attributed to moderately massive
(8--10 $M_{\odot}$) stars, ejecting their nucleosynthesis
contribution after some delay compared to the typical products of
very massive objects; this was adopted as a possible
explanation for the delay in the appearance of Eu and Ba with
respect to oxygen and iron (Travaglio et al.~1999). Making use
of this same separation of massive star yields in two groups with
different time scales for enrichment, we might improve our
suggestions on the origin of Cu and Zn, trying to disentangle
their $r$-process contribution from the rest. Actually, if the
$r$-process becomes well mixed in the Galaxy only after [Fe/H]
has reached the value $-$2.5 or so, we can tentatively interpret
the scattered Cu and Zn abundances in very metal-poor stars as
an indication that these last were born out of clouds selectively
contaminated by different supernova types, sometimes carrying
the signature of the $r$-process, sometimes that of NSE or
other explosive nucleosynthesis phenomena, in a poorly mixed
early Galaxy.

On the above hypothesis, a deeper insight into the contributions
from massive stars can in principle be obtained by inspection of a
few ultra-metal poor (UMP) stars like HD 115444 (Westin et
al.~2000), HD 122563 (Westin et al.~2000; this work), BD+173248
(Cowan et al.~2002), and CS 31082-001 (Hill et al.~2002),
provided Cu and/or Zn abundances, and their ratios to Ba or Eu
are available (note, again, that we take all the quantitative
information from observed data, so that the role of chemical
models is only to meld the ingredients, in order to see if
they form a self-consistent picture!). In particular, taking
the extremely $r$-process enriched ($\sim 40$ times times solar)
CS 31082-001 as an example, we can consider its abundances
as due to a virtually 'pure' $r$-process (i.e. the contribution by any
other nucleosynthesis mechanism is hidden by the dominant one from
the $r$-process). Knowing that the $r$-process accounts
for 20\% of Solar System Ba (Travaglio et al.~1999; Arlandini et
al.~1999), one might be tempted to derive directly the $r$-process
contribution to Zn (the abundance of Cu is not available for this
star). From this procedure one gets $\sim$ 2\% for
$r$-process Zn. Applying the same idea to Cu (e.g. in HD 115444)
one would obtain an $r$-process fraction for Cu $\sim$2\%, which
would in this case be an upper limit, as the star is not so
extremely $r$-process rich to exclude other contributions.
Unfortunately, a simple scheme like the one depicted, which works
fine for heavy $r$-elements, must be taken with great care for
lighter elements ($A < $140). Indeed, available abundance
observations in population II stars, especially of the UMP class,
have gradually introduced the idea that a unique $r$-process mechanism,
valid for both heavy and light $r$-elements does not exist (see
recent data in Sneden et al.~2000; Westin et al. 2000). Previous
sets of stellar abundances in the halo, pointing to the same
difficulties, and hints from isotopic anomalies related to
radioactive nuclei in the solar system, induced Wasserburg et al.
(1996) to suggest a galactic enrichment model in which the
$r$-process is at least bi-modal, with different supernova types
responsible for heavy ($A >$ 140) and light ($A <$ 140)
$r$-nuclei.  This new scenario, and especially the observational
material on which it is based, now prevents any detailed analysis of
the $r$-process contribution to Cu and Zn. We can only suggest
that a multiplicity of $r$-process components is expected to
add its effects to the scatter at low metallicity.

On the contrary, the general indications about primary massive
star yields can receive some independent confirmation from
a deeper look at the abundance distribution of individual UMP
stars. A recent study presented by Travaglio, Gallino, \&
Arnone~(2002), on the galactic chemical evolution of Sr, Y, and
Zr, showed that the abundances of these elements in UMP stars
suggests the existence of a primary neutron capture process in massive
stars (not the classical $r$-process). As discussed by Travaglio
et al.~(2002), this should also give a contribution of
$\sim$20\%-25\% to the Solar System composition of Sr, Y, and Zr.
Taking into account what we have discussed above, and using a
simple procedure similar to the one described before to derive
the $r$-fraction of Cu and Zn, we get what follows. We consider
HD122563 as a typical star enriched by the neutron-capture
processes occurring in advanced massive star evolution
('$n$-processes'). This implies ratios [(Sr, Y, Zr) to (Ba, Eu)] much
higher than average at those metallicities (see Travaglio et
al.~2002 for details). From this we can derive a primary
contribution to Cu and Zn by the $n$-process of $\sim$ 7\% and
$\sim$ 30\%, respectively (that is in fairly good agreement with the
contribution from massive stars derived in a completely different
way in the previous Section).

In conclusion, the general contribution from massive stars seems
to be roughly fixed from the database itself, despite the absence
of secure predictions from nucleosynthesis models. It can come
either from explosive Si burning, or NSE coinditions, or from the
relatively fast neutron captures called '$n$-process'. Instead,
due to the non-uniqueness of the $r$-process, we cannot specify
its relevance, but only suggest that the scatter at low metallicity
may be further amplified by the effects of different admixtures
of at least two $r$-process contributions and by other explosive
phenomena in supernovae.

\section{Conclusions}

In this paper we presented a large sample study of Cu and
Zn in metal-poor stars, after verifying the population of the
sample objects through estimates of their kinematic parameters.
In the absence of clear indications on the stellar
origin of Cu and Zn from the present status of nucleosynthesis
theories, we tried to derive suggestions on the relevant
mechanisms from an inspection of the observed data and from the
evolutionary trends expected for the outcomes of various nuclear
processes. In this way we inferred that several nucleosynthesis
phenomena are involved, suggesting that about 25\% of Cu is produced by
secondary phenomena in massive stars, and only 7-8\%  is due
to primary phenomena in the same environment (either explosive
or from a primary $n$-process). The bulk of Cu abundance (at least 62
-- 65\%) should be instead contributed on long time scales by
type Ia supernovae, in agreement with suggestions by Matteucci
et al. (1993). For Zn, its trend with respect to iron implies
similar percentage yields for the two elements: 1/3 from primary
processes in massive stars and 2/3 from type Ia supernovae.
These rough indications were shown to roughly account for the Cu and
Zn enrichment in the Galaxy. This approach was however found to be
too schematic for interpreting the details of the database,
including the scatter at very low metallicity. We argued that this
last might be due to poorly mixed different contributions from
massive stars in a non-homogeneous early stage (perhaps also
complicated to the non-uniqueness of the $r$-process). We also
concluded that, while a simple separation of the yields for Cu
and Zn in primary and secondary mechanisms, and in long-lived
and short-lived parents, is roughly possible, and gives a
preliminary guideline for future nucleosynthesis models, this
cannot be pushed too far, and we found no reliable tool for
disentangling the contributions of the two $r$-process
mechanisms from the rest. A more quantitative analysis must
therefore wait for a clarification of the underlying physical
processes in evolved stars.

\begin{acknowledgements}
We gratefully thank the suggestions contained in a detailed, very
collaborative referee report by Chris Sneden, whose contribution
greatly improved the presentation of our material. C.T thanks
Max-Planck-Institut f\"ur Astrophysik (Munich) for the support to
this work. M.B. thanks support from the MURST project Cofin2000
'Stellar Observables of Cosmological Relevance'.
\end{acknowledgements}

{}
\end{document}